\documentclass[twocolumn,twocolappendix]{aastex631}

\newcommand{\kms}{\mbox{\rm km s$^{-1}$}}

\newcommand{\NAOJ}{\affil{National Astronomical Observatory of Japan, 2-21-1 Osawa, Mitaka, Tokyo, 181-8588, Japan}}
\newcommand{\Shizuoka}{\affil{Faculty of Global Interdisciplinary Science and Innovation, Shizuoka University, 836 Ohya, Suruga-ku, Shizuoka 422-8529, Japan}}
\newcommand{\kogakuin}{\affil{Division of Liberal Arts, Kogakuin University, 2665-1 Nakano-cho, Hachioji, Tokyo 192-0015, Japan}}
\newcommand{\Shunan}{\affil{Faculty of Information Science, Shunan University, 843-4-2 Gakuendai, Shunan, Yamaguchi 745-8566, Japan}}
\newcommand{\shibaura}{\affil{College of Systems Engineering and Science, Shibaura Institute of Technology, 307 Fukasaku, Minuma-ku, Saitama City, Saitama 337-8570, Japan}}

\newcommand{\ithems}{\affil{Interdisciplinary Theoretical \& Mathematical Science Center (iTHEMS), RIKEN, 2-1 Hirosawa, Wako, Saitama 351-0198, Japan}}
\newcommand{\ipmu}{\affil{Kavli Institute for the Physics and Mathematics of the Universe (WPI), UTIAS, The University of Tokyo, 5-1-5 Kashiwanoha, Kashiwa, Chiba 277-8583, Japan}}
\newcommand{\kagawa}{\affil{Faculty of Education, Kagawa University, Saiwai-cho 1-1, Takamatsu, Kagawa 760-8522, Japan}}
\newcommand{\oec}{\affil{Research Center for Physics and Mathematics, Osaka Electro-Communication University, 18-8 Hatsucho, Neyagawa, Osaka 572-8530, Japan}}
\newcommand{\sokendai}{\affil{Astronomical Science Program, Graduate Institute for Advanced Studies, SOKENDAI, 2-21-1 Osawa, Mitaka, Tokyo 181-1855, Japan}}

\shorttitle{An AGN in the Antennae galaxies ?}
\shortauthors{Komugi et al.}

\begin{document}

\title{An AGN in the Antennae galaxies ?}

\correspondingauthor{Shinya Komugi}
\email{skomugi@cc.kogakuin.ac.jp}
\author[0009-0007-2493-0973]{Shinya Komugi}\kogakuin
\author[0000-0002-2501-9328]{Toshiki Saito}\Shizuoka
\author[0000-0003-2475-7983]{Tomonari Michiyama}\Shunan
\author[0000-0002-7272-1136]{Yoshiyuki Inoue}\shibaura\ithems\ipmu
\author[0000-0002-6939-0372]{Kouichiro Nakanishi}\NAOJ
\author[0000-0002-2062-1600]{Kazuki Tokuda}\kagawa
\author[0000-0002-8868-1255]{Fumiya Maeda}\oec
\author[0009-0008-9573-4766]{Yuzuki Nagashima}\sokendai

\begin{abstract}
Time variability is a strong probe of energetic phenomena which occur at small spatial scales, like Active Galactic Nuclei (AGN).
We use ALMA observations at 100 GHz executed over a period of $\sim$2.5 months to look for time variability in the Antennae galaxies, a prototypical early stage merger galaxy pair, for which there are no previous signatures of an AGN in the optical, infrared or X-ray. Most 100 GHz detections in the Antennae are spatially extended and associated with star forming regions, but two sources in the southern galaxy NGC\ 4039 are compact.  One of these compact sources, S3, is offset by $\sim 1^{\prime \prime}$ in the northeast direction from the stellar peak of NGC\ 4039, and marginally resolved at 10 parsec resolution.  The other source, S4, is co-spatial with the stellar peak of NGC\ 4039 and unresolved even at a resolution of $\sim 4$ parsec. We examine the time variability of these two sources using their power spectrum.  We find that S4 varies with a characteristic timescale of $13\pm 3$ days, indicating that the phenomena responsible for the 100 GHz emission is smaller than 0.01 parsecs.  By comparing the observed flux of the two sources with various candidate origins, we show that while S3 can be explained either by a young massive stellar cluster or an AGN, S4 is likely to be an AGN that is possibly Compton-thick.

\end{abstract}



\section{Introduction} \label{sec:intro}






Interactions and mergers with other galaxies is believed to be an important pathway during galaxy evolution, as it may trigger an Active Galactic Nuclei (AGN).  X-ray observations have shown that post-merger galaxies host a larger fraction of AGN compared to a control sample \citep{Li2023}.  It is not known, however, when in the merging process the AGN is actually triggered.

The Antennae galaxies (NGC~4038/4039) lie at a distance of 22~Mpc
\citep{Schweizer08}, and is one of the nearest gas-rich merging galaxy pairs.  They are also one of the youngest mergers, thought to be only $\sim 300$ Myr since their first passage \citep{Renaud2009}.
Previous studies have not found any signs of an AGN in this system. Chandra X-ray observations \citep{Zezas2002a, Zezas2002b} detected more than 40 sources in this system, most of which could be explained by X-ray binaries and thermal processes. Infrared spectra from Keck \citep{Gilbert2000} and Spitzer \citep{Brandl2009} do not find any AGN signatures, either. Although the nucleus of NGC~4039 indicates a high [NeIII]/[NeII] ratio and bright molecular $\mathrm{H_2}$ emission, the typical AGN indicator [NeV] was not detected, and furthermore, the spectrum was consistent with a starburst.  However, another promising avenue for searching for AGN is to look for time variability, which would indicate activity at small spatial scales.
\cite{He2022} used Atacama Large Millimeter/submillimeter Array (ALMA) observations to study young massive stellar clusters, and in the course of doing so, found that flux from two observations 2 years apart do not match in two sources near the nucleus of NGC~4039, the southern galaxy.

In this paper we use high resolution ALMA observations at 100~GHz (3~mm) conducted over  a $\sim$2.5 month period to search for time variability in millimeter flux.  In Section 2 we explain the observations and the data reduction procedure.  In Section 3 we compare the ALMA data of two sources in NGC~4039 with images at other wavelengths, present their light curves, and discuss their time variability.  In Section 4 we explore the nature of the time variation and possible mechanisms.






\begin{figure*}[t!]
\begin{center}
\includegraphics[width=18.5cm]{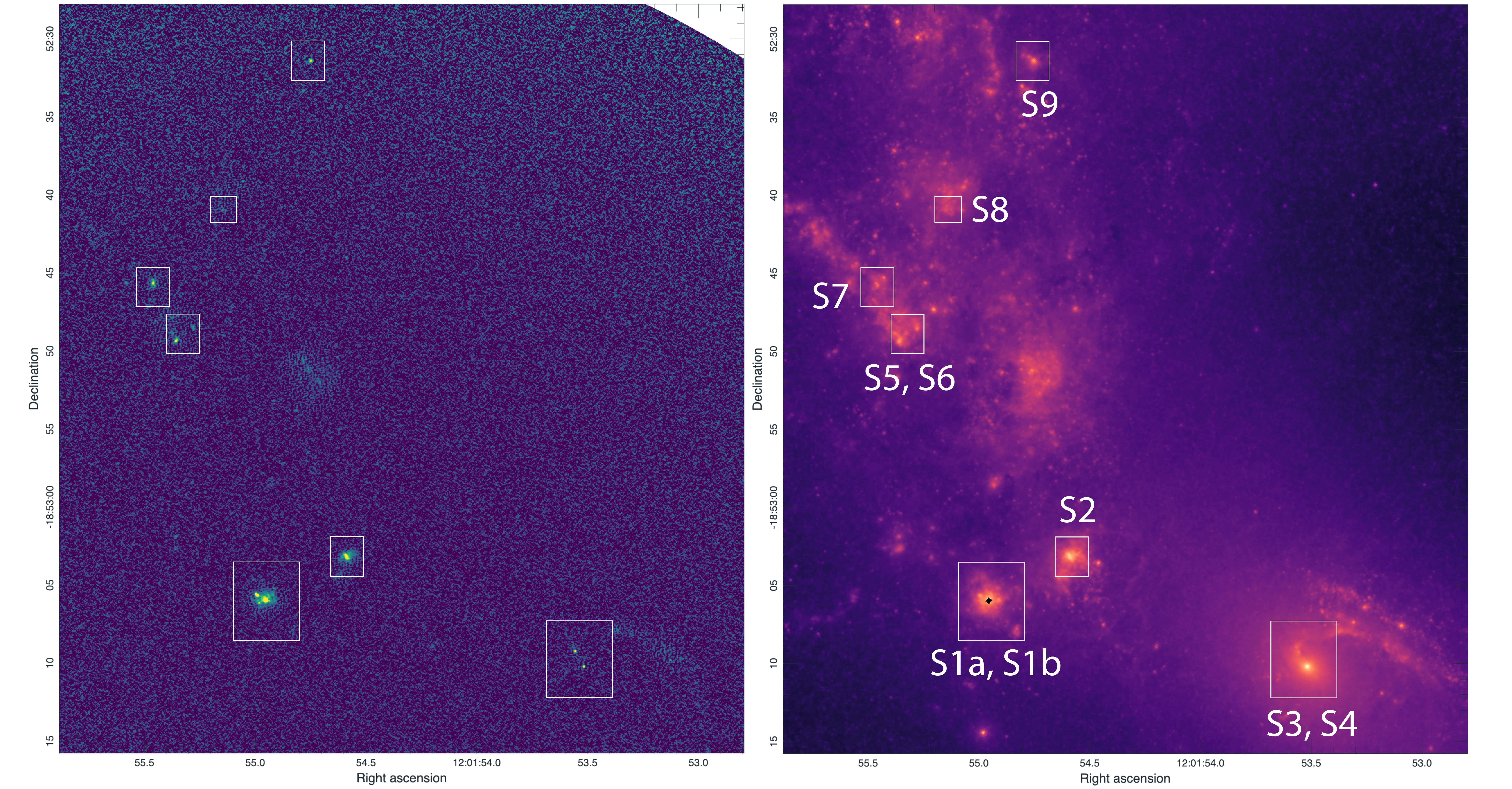}
\end{center}
\caption{
The combined ALMA 3~mm continuum image (left) and {\it JWST} NIRCam $3.35~\mathrm{\mu m}$ (right) image of the Antennae (only the overlap region and NGC~4038 are shown).  Individual regions are shown in rectangles.}\label{map}
\end{figure*}

\section{Observations and Data Processing} \label{sec:obs}

The data were obtained at the ALMA 12m array as one of the Cycle~9 observatory filler projects\footnote{\url{https://almascience.nao.ac.jp/news/alma-announces-observatory-projects-for-configurations-8-and-9}} (Project Code: 2022.A.00032.S).  The observations were conducted between 2023 September 5th and November 23rd under good weather conditions for Band~3 (PWV $=$ 0.33$-$5.45~mm with a median of 1.385~mm). The total number of executions (Execution Block, EB) was 52 resulting in a total on-source time of 40.3~hours (See Table \ref{data} in the Appendix). We configured four spectral windows (spw) each with 1.875~GHz bandwidth ($\gtrsim$ 4900~\kms). The central frequency of each spw is tuned to 114.64195~GHz, 112.77718~GHz, 102.63664~GHz, and 100.77188~GHz with the spectral resolution of 1.129~MHz (2.95~\kms), 1.953~MHz (5.19~\kms), 31.250~MHz (91.3~\kms), and 31.250~MHz (93.0~\kms), respectively.  The 112.7 GHz spw includes two CN ($N$ $=$ 1--0) rotational transitions ($J$ $=$ 3/2--1/2 and 1/2--1/2), but with no significant detection. The 114.6~GHz spw includes the $^{12}\mathrm{CO} (J$=1--0) emission.  
Details of the CO emission will be published in a forthcoming paper (Saito et al., in preparation).
The calibrated data (Measurement Sets) were provided by the observatory, and imaged using CASA version 6.5.2--26. In section \ref{sec:multi} we show the CO emission image created using the PHANGS-ALMA pipeline \citep{Leroy21}. For the continuum, we use three imaging versions of the data explained below, for different purposes. 

\subsection{combined image}
The \textit{combined} image uses all EBs to robustly measure the flux and spatial extent of the millimeter emission. All spws were used, with any detected emission lines masked out.  We used clean masks provided by the observatory as part of the QA2 assessment, which were created from auto-masking, and cleaned to a threshold of $1~\mathrm{\mu Jy}$ (0.5$\sigma$).  Imaging was done with the Briggs weighting (robust parameter=0.5), which resulted in a beam size of $0^{\prime \prime}.10\times 0^{\prime \prime}.087$
 with a position angle of $-75.6$ deg., corresponding to a projected scale of $11\times 9.3$ pc.  The r.m.s. noise level was $2.1~\mathrm{\mu Jy}$. In Figure \ref{map} we show the combined continuum image.

\begin{figure*}[t!]
\begin{center}
\includegraphics[width=18cm]{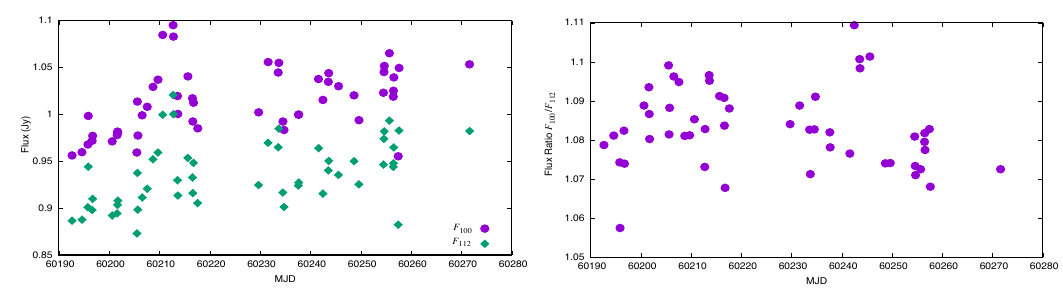}
\end{center}
\caption{Flux variation of the phase calibrator, J1215-1731, over the observing run. The horizontal \textbf{axis} is the Modified Julian Day (MJD) of the observation. Left panel : flux variation at 100.7 GHz ($F_\mathrm{100}$) and 112.7 GHz ($F_\mathrm{112}$).  Right : Flux ratio between two frequencies. The ratio has an average of $1.08\pm 0.01$, i.e., stable to $\sim 1\%$}\label{phasecal}
\label{fig:scatter}
\end{figure*}

\subsection{uniform image}
The \textit{uniform} image uses all EBs but with larger weighting on longer array baselines to attain the highest angular resolution. We use the uniform image to obtain upper limits on the size of point sources. We use only the three continuum spw centered at 100.7, 102.6 and 112.7 GHz.  
The images were created by using the CASA task \textit{tclean} with \textit{weighting=uniform}, and user defined masks that were centered on known sources S1a, S1b, S2, S3, S4 and S7 \citep{He2022} with a radius of 0.1 arcsec.  Cleaning was stopped at 1$\times$ the r.m.s. noise level of $17~\mathrm{\mu Jy}$.  The resulting beam size was $0^{\prime \prime}.045\times 0^{\prime \prime}.039$
 with a position angle of 86.8 deg., corresponding to a projected scale of $4.7\times 4.1$ pc.

\subsection{individual image}
Using the same spws as the uniform image, we imaged all EBs separately to search for time variability.
The imaging was done with the Briggs weighting (robust parameter=0.5), same as the combined image.
All EBs were imaged using a combination of automatic masking and the mask used in the \textit{combined} image, and cleaned down to the 1$\times$ r.m.s. noise level, where the noise was measured for regions where the primary beam (PB) response was greater than 50\%.  The pixel size of each image was set to be $\frac{1}{3}$ of the synthesized beam size.  The beam size and r.m.s. noise level are tabulated in Table \ref{data} in the Appendix.


\subsection{Flux Calibration}\label{fluxcal}
Amplitude calibration was done using the bandpass calibrator given in Table \ref{data} in the Appendix.  The absolute calibration accuracy is $\sim 5\%$ or better (see ALMA Technical Handbook 10.2.6, \citealp{Guzman19,Farren21}).
Throughout the observing run, QSO J1215-1731 was used as the phase calibrator.  To check for the stability of flux calibration, we fit the visibility of the phase calibrator with a point source model for each EB using CASA task \textit{uvmodelfit}.  Figure \ref{phasecal} (left panel) shows the phase calibrator flux over the observing run, for the 100.7 GHz ($F_\mathrm{100}$) and 112.7 GHz ($F_\mathrm{112}$) spws. Results from the spw centered at 102.7 GHz were indistinguishable from 100.7 GHz.  
Uncertainties due to point source fitting were negligible.  Flux in both spws follow the same trend, with peak-to-peak variation of $\sim 10\%$.  Overall, we find $F_\mathrm{100}=1.01\pm 0.03\ \mathrm{Jy}$ and $F_\mathrm{112}=0.94\pm 0.03\ \mathrm{Jy}$, i.e., the uncertainties are $~3\%$.  This includes uncertainties coming from absolute flux calibration, and also the intrinsic flux variation of the calibrator. The flux ratio $\frac{F_\mathrm{100}}{F_\mathrm{112}}$ (Figure \ref{phasecal} right panel) is stable to within $1\%$, indicating the stability of relative calibration between two different spws.
The uncertainties are smaller than the nominal absolute flux calibration accuracy of ALMA ($5\%$).  Therefore, we apply a conservative absolute flux uncertainty of $5\%$ for the analyses presented in this paper.  In any case, the flux uncertainty coming from aperture photometry, discussed hereafter, is a larger source of error for most sources.

\subsection{Photometry}\label{photometry}
Since the observations span nearly 3 months, many of the executions were observed with different antenna configurations and consequently have different beam sizes.
The beam size for each execution is listed in Table \ref{data} in the appendix.
The beam size steadily increases toward the later runs, up to a beam diameter of $~0^{\prime \prime}.3$.
In order to avoid introducing systematics in the analyses, we need to measure the flux of regions using a uniform aperture.
As a reasonable balance between trying to capture the flux in the outskirts of point-like structures observed with large beams, and trying to mitigate the effect of sidelobes in images with smaller beams, we choose an aperture diameter of $0^{\prime \prime}.5$ throughout all executions to measure the flux in individual images.  For total flux measurements in the combined image, we use larger apertures to enclose the bulk flux in extended sources.  For all apertures, the maximum recoverable scale (MRS) of the observations is larger than the aperture, so we do not expect flux to be significantly resolved out. Aperture diameters are listed in Table \ref{sources} and MRS are listed in Table \ref{data} in the appendix. Uncertainties coming from aperture photometry were estimated by multiplying the noise level by the square root of the number of resolution elements in the aperture, and range from 1\% (source S1) to 20\% (source S8).  These were added in quadrature with the absolute flux uncertainty of 5\% explained in section \ref{fluxcal} to give the uncertainties listed in Table \ref{sources}.

\subsection{Ancillary Data}\label{ancillary}
JWST NIRCAM and MIRI images at $1.15~\mathrm{\mu m}$, $3.35~\mathrm{\mu m}$ and $7.70~\mathrm{\mu m}$, discussed in section \ref{sec:image}, were obtained from the Mikulski Archive for Space Telescopes (MAST) at the Space Telescope Science Institute. The images are fully calibrated, pipeline stage 3 products. The specific observations analyzed can be accessed via \dataset[doi:10.17909/zar5-cy27] {https://doi.org/10.17909/zar5-cy27}.

\begin{deluxetable*}{lccccc}[ht!]
\tablecaption{3 mm flux of Continuum sources \label{sources}}
\tablewidth{0pt}
\tablehead{
\colhead{} & R.A. & Dec. & Aperture ($^{\prime \prime}$)& 3~mm Flux (mJy) & Notes 
}
\startdata
S1a &  12:01:54.952 & -18:53:05.98 & 0.8 & $1.937\pm 0.099$ & Extended, multiple components \\
S1b &  12:01:54.989 & -18:53:05.63 & 0.5 & $0.706\pm 0.037$ & Extended, multiple components \\
S2  &  12:01:54.591 & -18:53:03.06 & 0.8 & $0.906\pm 0.049$ & Extended southwest \\
S3  &  12:01:53.555 & -18:53:09.22 & 0.5 & $0.153\pm 0.013$ & Point source, slightly extended \\
S4  &  12:01:53.516 & -18:53:10.20 & 0.5 & $0.157\pm 0.014$ & Point source \\
S5  &  12:01:55.357 & -18:52:49.34 & 0.5 & $0.265\pm 0.017$ & Extended\\
S6  &  12:01:55.280 & -18:52:48.43 & 0.5 & $0.082\pm 0.012$ & Extended south \\
S7  &  12:01:55.459 & -18:52:45.65 & 0.7 & $0.415\pm 0.026$ & Extended, multiple components \\
S8  &  12:01:55.146 & -18:52:40.97 & 0.3 & $0.035\pm 0.007$ & Slightly extended \\
S9  &  12:01:54.747 & -18:52:31.39 & 0.5 & $0.212\pm 0.028$ & Extended \\
\enddata
\tablecomments{Region S9 is outside of PB$>50\%$ coverage.  Noise for S9 is estimated to be 4.8 uJy/b, which we use for estimation of its flux uncertainty.}
\end{deluxetable*}

\section{Millimeter continuum in the overlap and NGC~4039}\label{sec:image}
\subsection{Morphology of continuum sources}
\begin{figure*}[t!]
\begin{center}
\includegraphics[width=8.5cm]{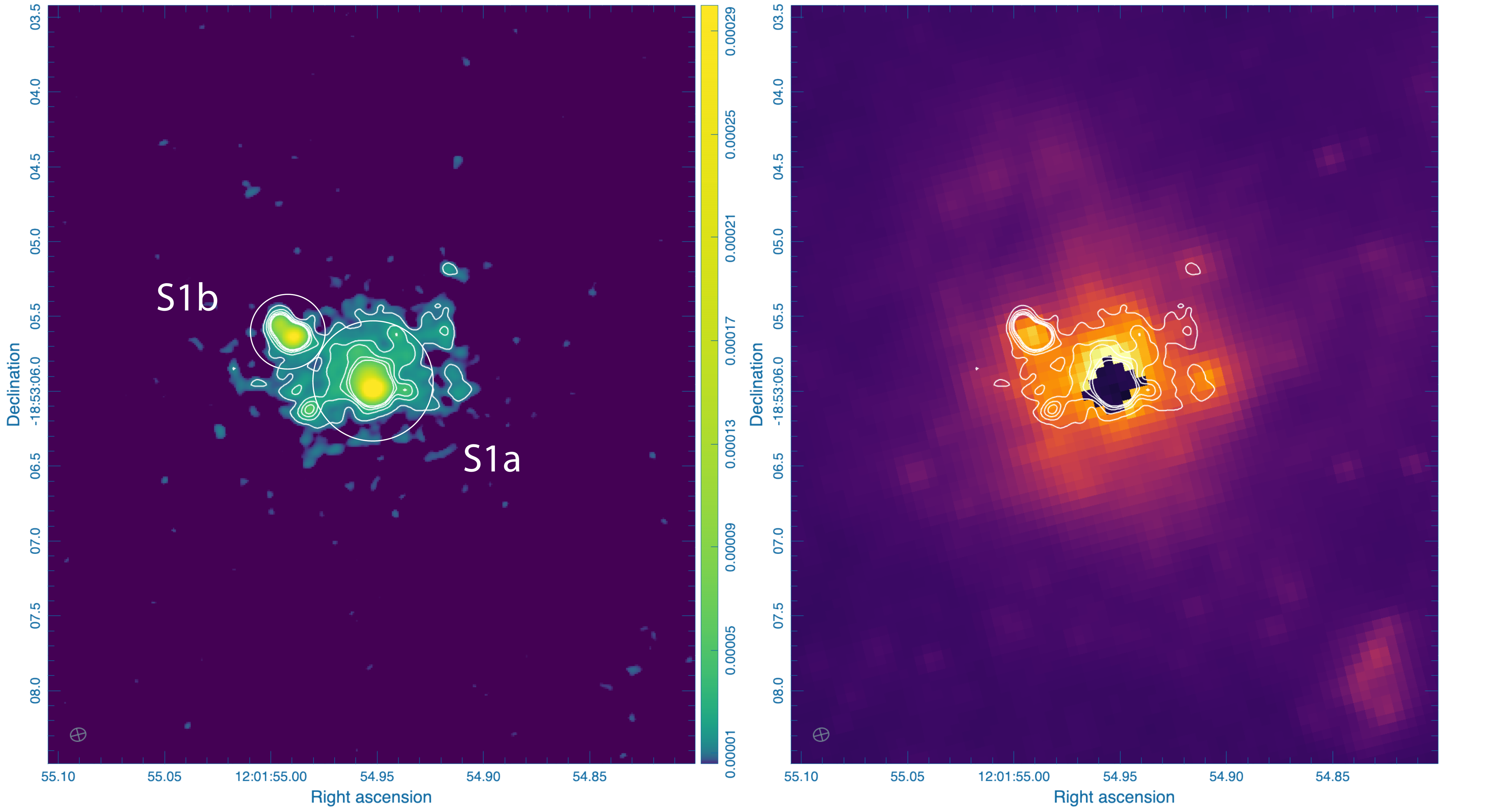}
\includegraphics[width=8.5cm]{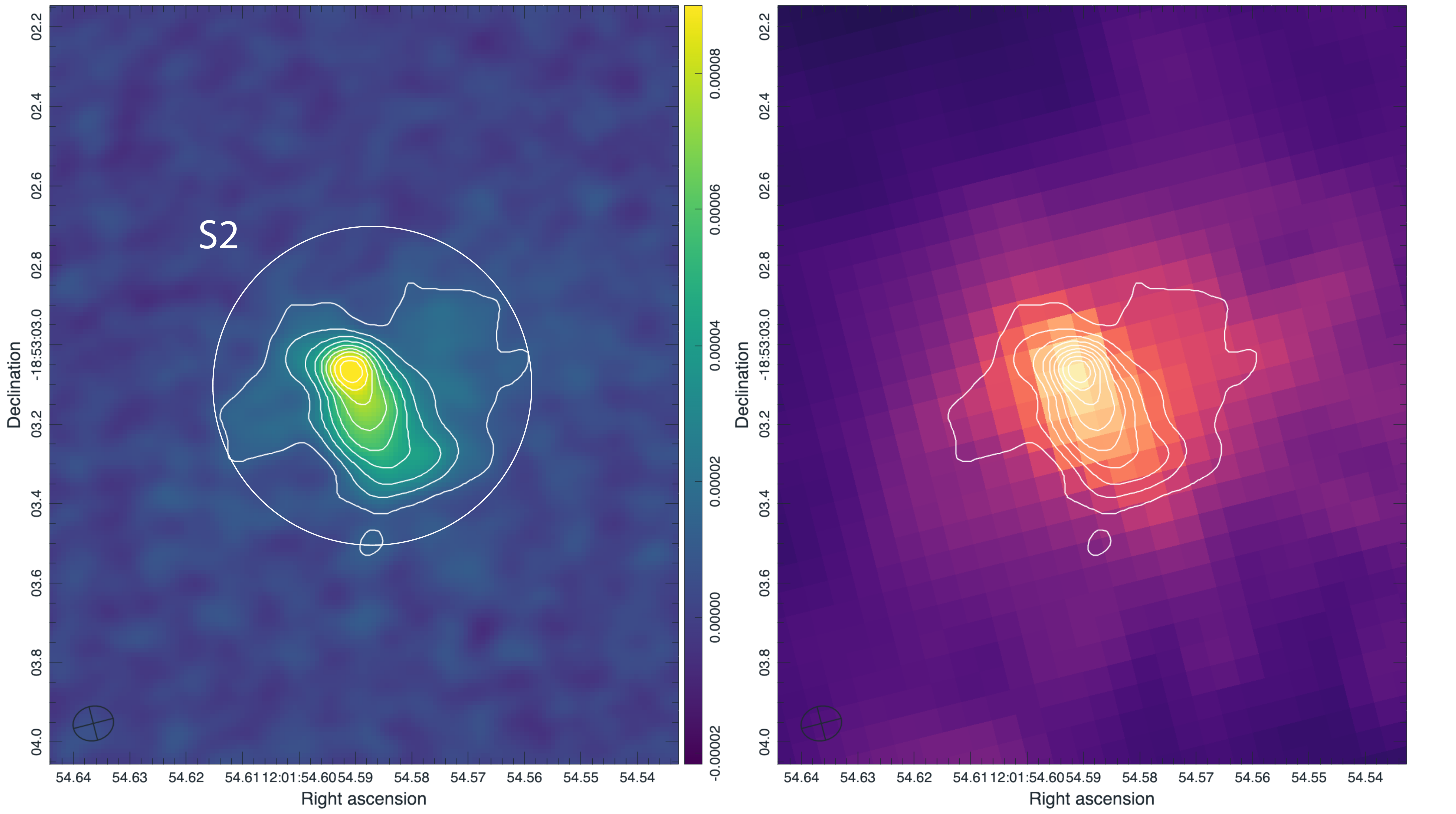}\\
\includegraphics[width=8.5cm]{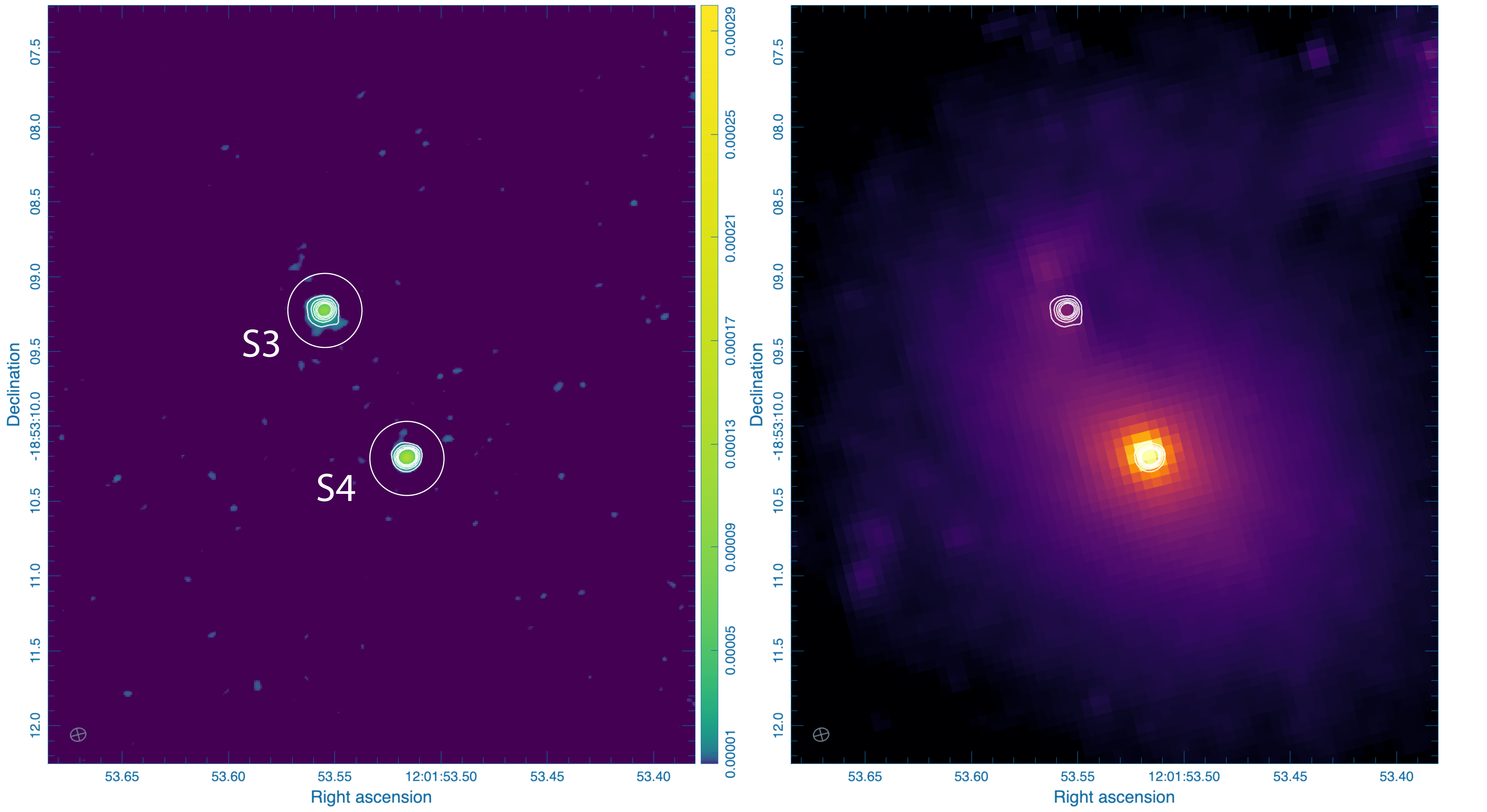}
\includegraphics[width=8.5cm]{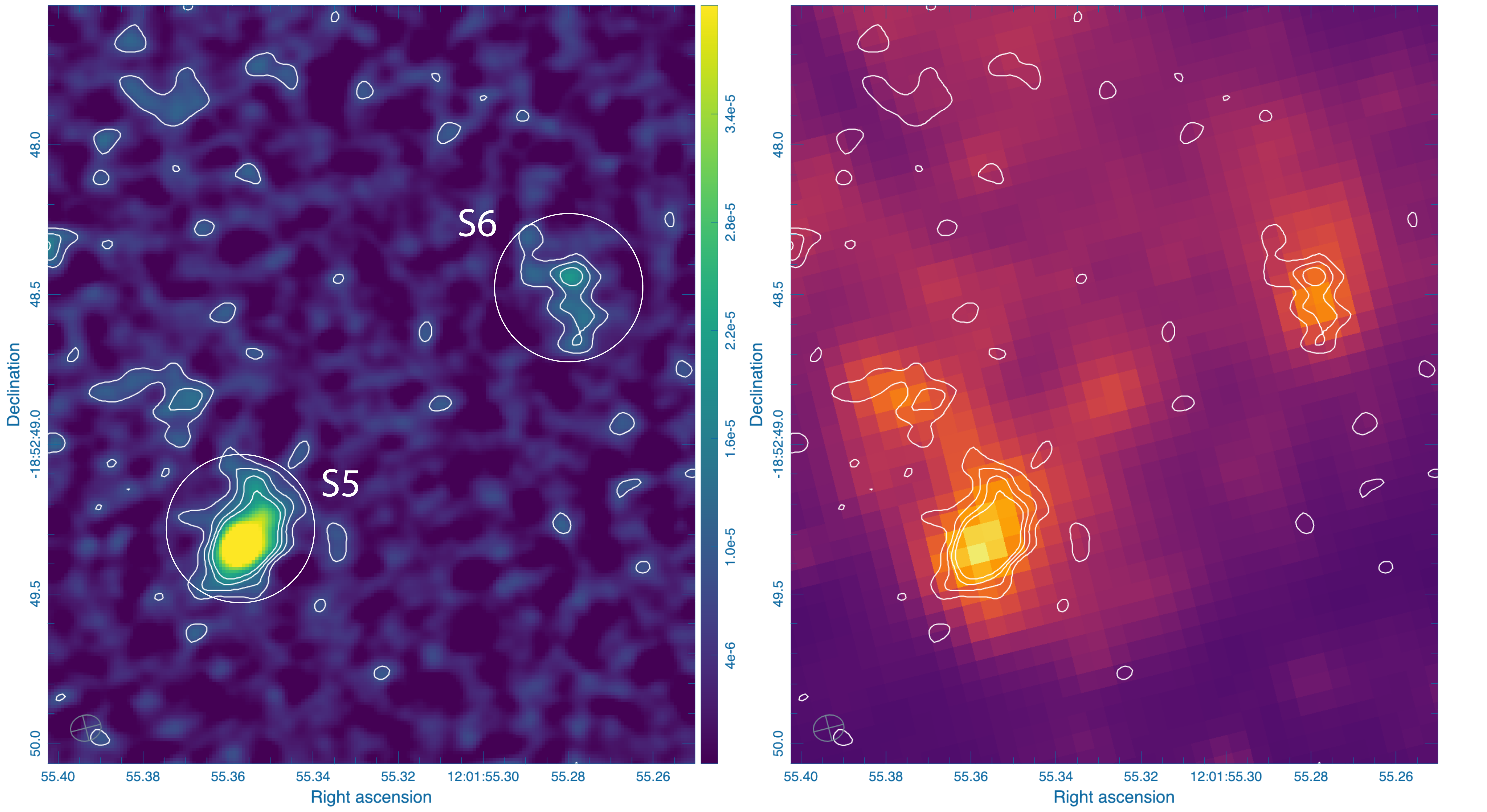}\\
\includegraphics[width=8.5cm]{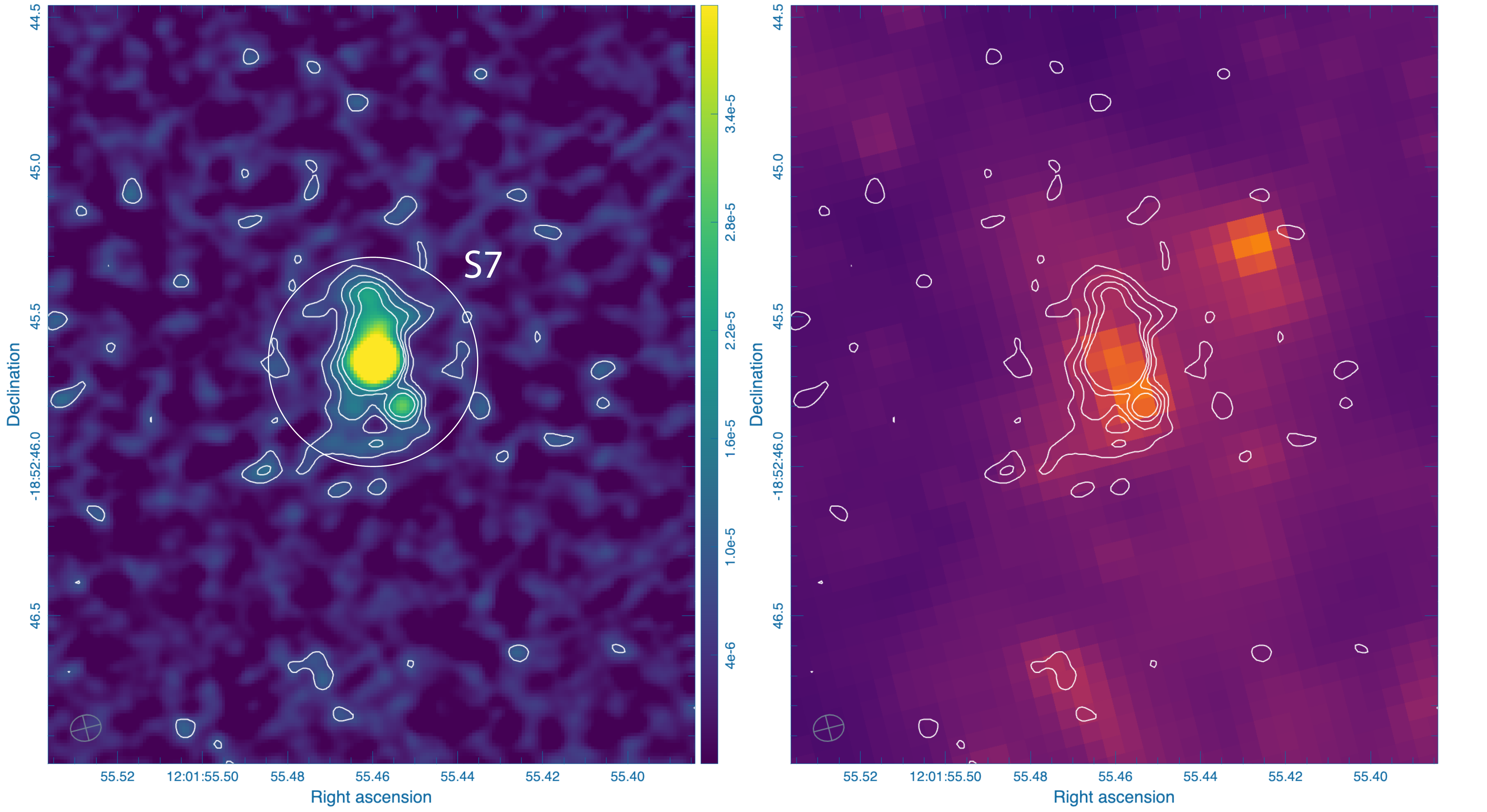}
\includegraphics[width=8.5cm]{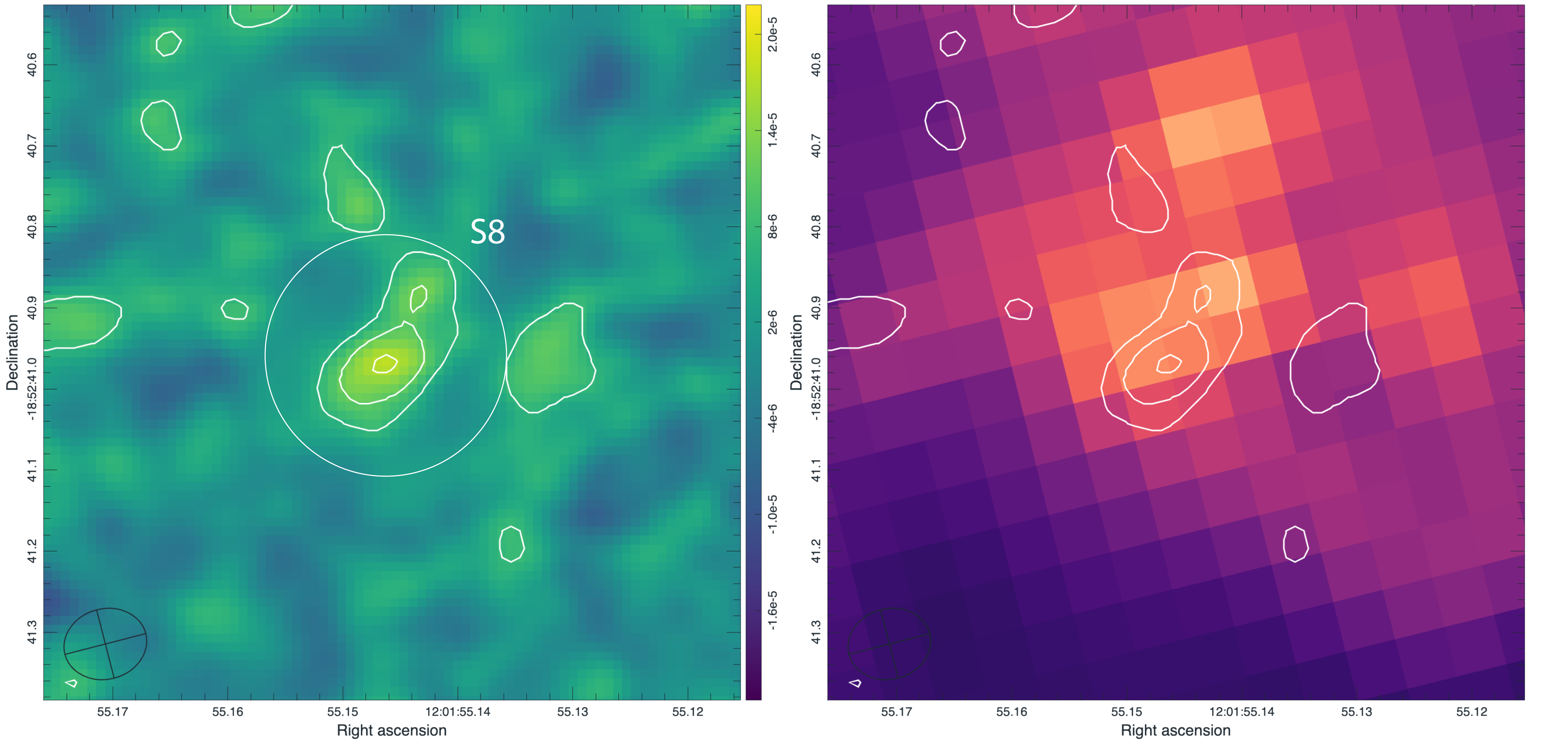}\\
\includegraphics[width=8.5cm]{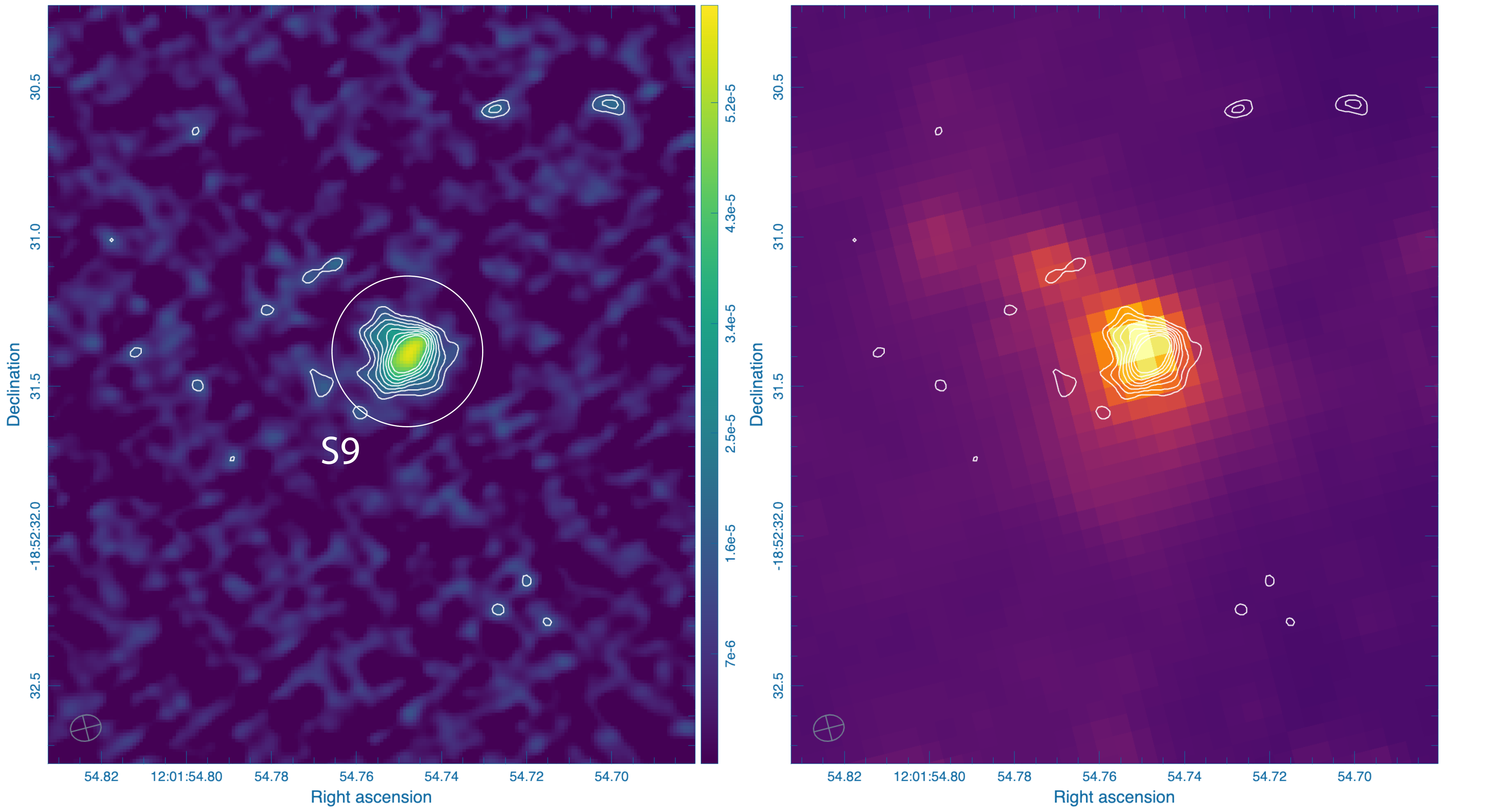}\\
\end{center}
\caption{Closeup of regions S1 to S9.  Left : Combined image at 3mm. Contours are drawn from 5$\sigma$ in increments of 5$\sigma$ for S1a, S1b, and S2. At 5, 10, 20 and 30 $\sigma$ for S3 and S4.  From 3$\sigma$ in increments of 2$\sigma$ for S5, S6, S7, S8, and S9. Aperture used to measure the flux is also shown in circles. All units are in Jy. Right : JWST NIRCAM $3.35\ \mathrm{\mu m}$ image. 
\label{regions}}
\end{figure*}

In Figure \ref{regions} we show cutouts of regions from the combined image that were already detected and measured in previous ALMA data \citep{He2022}, along with JWST NIRCam images at $3.35~\mathrm{\mu m}$ which indicate warm dust and polycyclic aromatic hydrocarbon (PAH) molecules heated mainly by star formation. Sources S1 to S9 from \cite{He2022} overlap with the images presented here.  
Figure \ref{map} clearly shows extended structures at 100~GHz that do not overlap with S1-S9, most notably the spiral structure in the northwest of S3/S4, and in the north of S2.  These also have corresponding structure at $3.35~\mathrm{\mu m}$.
A more exhaustive catalog of continuum sources detected in the combined image will be presented in a subsequent paper.  The measured 3~mm continuum flux of sources S1-S9 is listed in Table \ref{sources}.  All of these sources, except for S3 and S4, are extended and readily identifiable with star-forming regions in JWST images.
The similarity of the 3~mm and $3.35~\mathrm{\mu m}$ morphology suggest that
 the 3~mm continuum is dominated by thermal free-free emission, most likely powered by young massive stellar clusters \citep{He2022}, for S1, S2, and S5 to S9.  S3 and S4 stand out as conspicuous point-like sources. Hereafter, we focus on the nature of the two sources S3 and S4.

\subsection{Spatial Extent of S3 and S4}
We find that S3 is slightly extended, while S4 is consistent with a point source. The upper left panel in Figure \ref{s3-s4} show the contours of the residual after fitting and subtracting a point source (image-based 2-d Gaussian) 
 from S3 and S4.

S3 has residual emission extending towards the southwest and northeast.  In the uniform image, it is point-like, but a two-dimensional Gaussian fit yields an integrated flux of $83\pm 26~\mathrm{\mu Jy}$ with a FWHM of $0^{\prime \prime}.035 \times 0^{\prime \prime}.028$.  The flux is half of what is measured in the combined image (see Table 1). The loss of flux indicates that S3 is partly resolved out in the uniform image.

S4 does not have any significant residuals from the point source subtraction. A two-dimensional Gaussian fit to the uniform image of S4 gives a FWHM of $(0^{\prime \prime}.036\pm 0.004)\times (0^{\prime \prime}.047\pm 0.008)$, consistent with the uniform image beam size, and a flux of $117\pm 34~\mathrm{\mu Jy}$, within 2 sigma of the combined image.  S4 is therefore consistent with a point source at the resolution of 4 pc.

The stellar peak of NGC~4039, measured as the peak of JWST NIRCam image at $1.15~ \mathrm{\mu m}$, is (R.A., Dec.)=(12h01m53.5174s, -18d53m10.176s).  This is offset from the peak position of S4 by $0^{\prime \prime}.031$.  Given that the pixel size of NIRCam is $0^{\prime \prime}.031$ and this is only 1/3 of the ALMA beam size, we hereafter assume that S4 is associated with the stellar peak of NGC~4039.




\subsection{Spectral Index of S3 and S4}\label{sec:index}
The low and high frequency spectral windows of the ALMA observations are separated by $\sim$10\ GHz, which allow an estimation of the spectral index in the millimeter range, under the assumption that the spectral index does not change with time.  We created the combined image separately for the $100~ \mathrm{GHz}$, $102~ \mathrm{GHz}$ and $112~ \mathrm{GHz}$ spw in the same way as the combined image and measured the flux using the same aperture diameter of $0^{\prime \prime}.5$ diameter.  
For S3, the measured flux were $176\pm 20\ \mathrm{\mu Jy}$, $161\pm 20\ \mathrm{\mu Jy}$ and $177\pm 31\ \mathrm{\mu Jy}$ for 100~GHz, 102~GHz and 112~GHz, respectively.
For S4, the measured flux were $176\pm 20\ \mathrm{\mu Jy}$, $158\pm 20\ \mathrm{\mu Jy}$ and $194\pm 32\ \mathrm{\mu Jy}$ at 100~GHz, 102~GHz and 112~GHz, respectively.
The flux at different frequencies are consistent within their uncertainties, indicating a flat or shallow spectrum.
Low resolution ($\sim 1^{\prime \prime}$) observations \citep{Neff2000} at 4~cm and 6~cm
detected sources at positions consistent with S3 and S4.  In Table 5 of \cite{Neff2000}, the coordinates of the first and second row of Region 1:2 correspond to S4 and S3, respectively. 
Combining the millimeter flux with these VLA measurements, we estimate the cm-mm spectral slope (assuming $F_\nu \propto \nu^{\alpha}$) to be 
$\alpha^{\mathrm{cm-mm}}_\mathrm{S3}\sim -0.12\pm 0.03$ and $\alpha^{\mathrm{cm-mm}}_\mathrm{S4}\sim -0.33\pm 0.04$. These slopes should be considered crude underestimates and indicative at best, because there could be contamination from the surrounding regions in the low resolution VLA measurements, and also because the VLA observations were obtained at a different epoch. A possible time variability may affect the derived cm-mm spectral slope.  Although it is difficult to draw robust conclusions about the spectral shape given the narrow frequency range of the ALMA observations, the spectrum of S4 seems to become shallower in the millimeter compared to the centimeter range, while S3 does not show significant changes.

\subsection{Flux variation of S3 and S4}
Time variability of millimeter emission can give vital clues to the origin of sources. Incidentally, the two sources S3 and S4 are also the ones that showed inconsistent flux between two observation epochs in \cite{He2022}, hinting at a possibility of time variability.

\begin{figure*}[t!]
\begin{center}
\includegraphics[width=17cm]{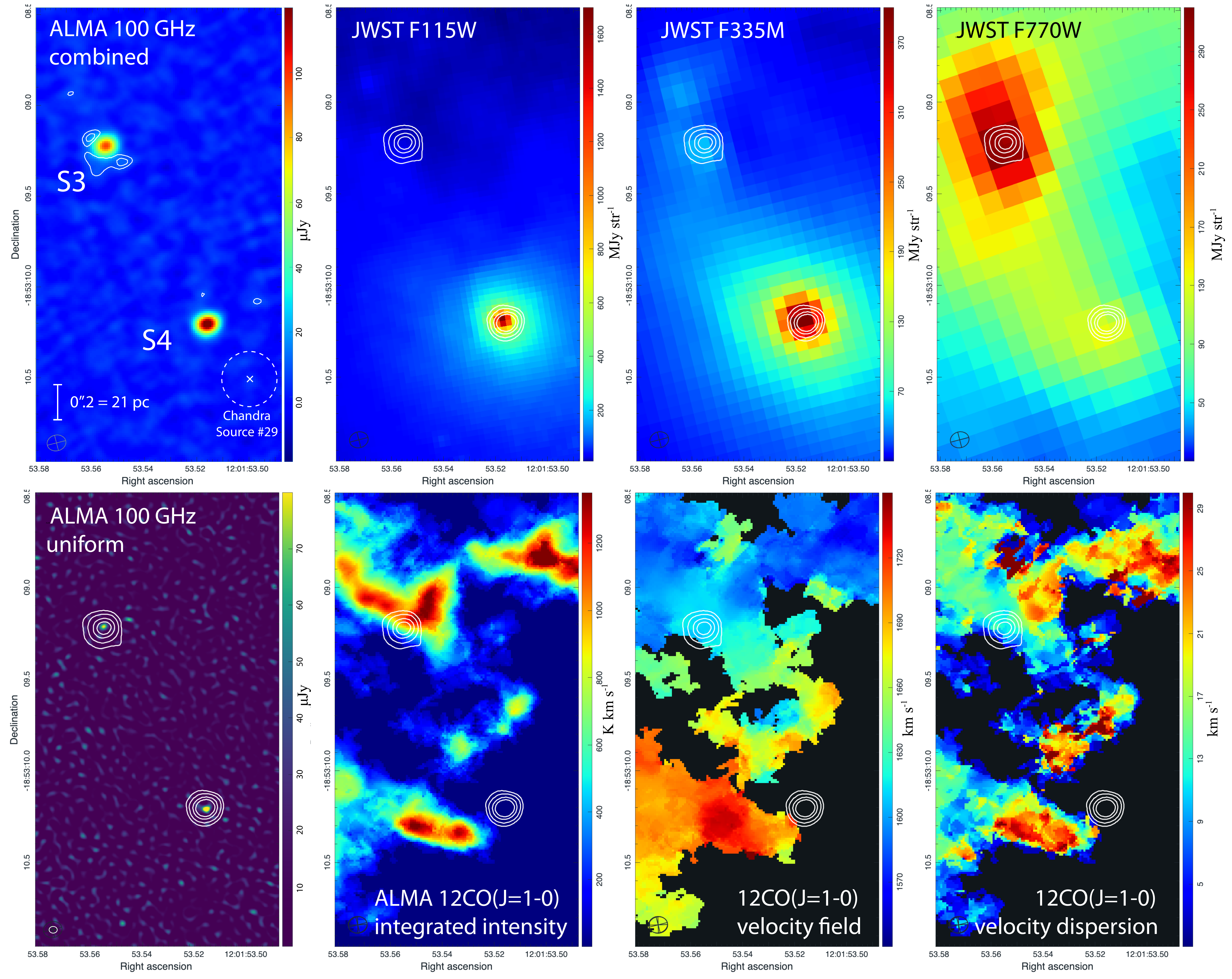}
\end{center}
\caption{Multi-wavelength view of S3 and S4.  From upper left to lower right: Combined ALMA image at 100~GHz ($\mathrm{\mu}$Jy), JWST NIRCam $1.15\ \mathrm{\mu m}$, JWST NIRCam $3.35\ \mathrm{\mu m}$, JWST MIRI $7.70\ \mathrm{\mu m}$ ($\mathrm{MJy\ str^{-1}}$), high resolution \textit{uniform} ALMA image at 100~GHz (Jy),  $^{12}\mathrm{CO}(J=1-0)$ integrated intensity ($\mathrm{K\ km\ s^{-1}}$), CO intensity weighted velocity (moment 1: $\mathrm{km\ s^{-1}}$), and CO velocity dispersion (moment 2: $\mathrm{km\ s^{-1}}$) image.  CO images have a beam size of $0^{\prime \prime}.10\times 0^{\prime \prime}.09$.
Contours in the upper left panel indicate the residual after point source subtraction, at 3, 4 and 5 $\sigma$. For all other panels,} contours are from the combined image, at 5, 10, 20 and 30 $\sigma$.  The white cross in the upper left panel indicates the position of the ULX source 29, and the dashed white circle indicates its uncertainty in position (see section \ref{sec:multi}).
\label{s3-s4}
\end{figure*}

\begin{figure*}[t!]
\begin{center}
\includegraphics[width=17cm]{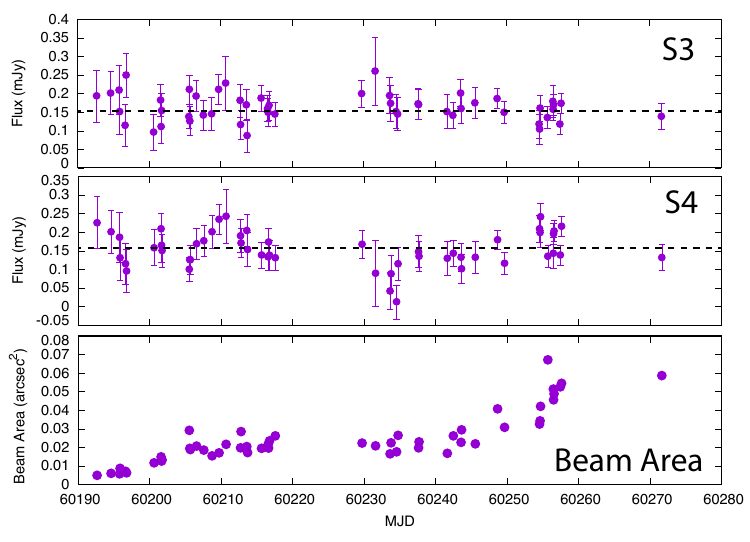}
\end{center}
\caption{Flux variation of S3 (top) and S4 (middle).  The fluctuation of the beam area is also shown (bottom).}
\label{lightcurve}
\end{figure*}

\begin{figure*}[t!]
\begin{center}
\includegraphics[width=8.5cm]{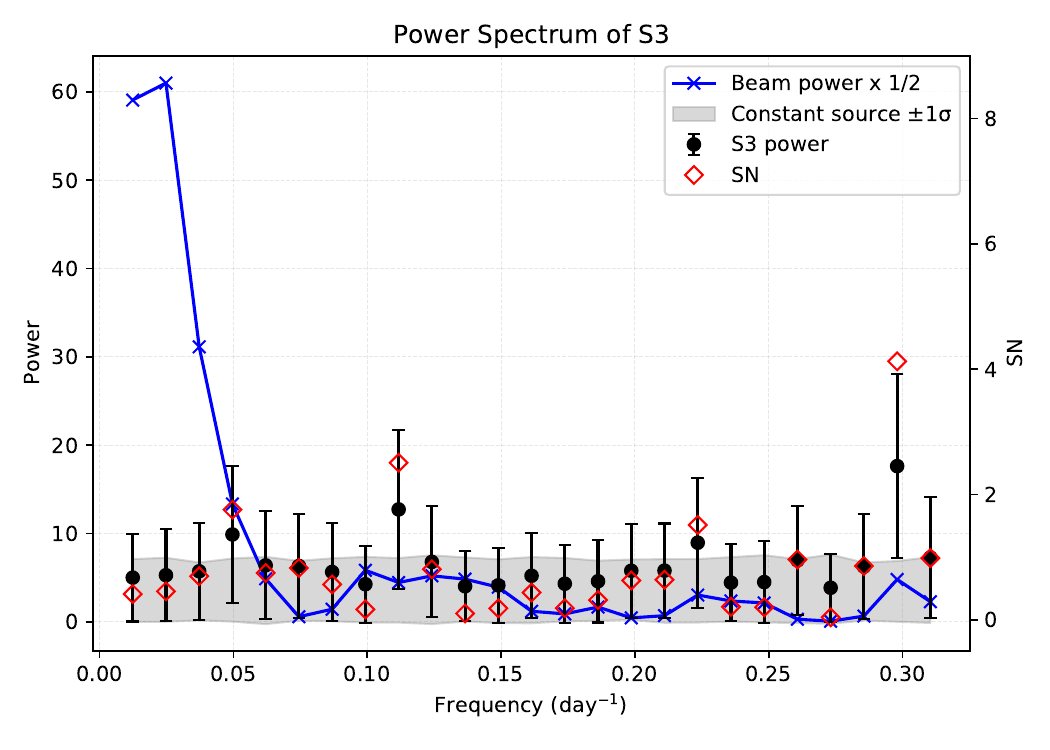}
\includegraphics[width=8.5cm]{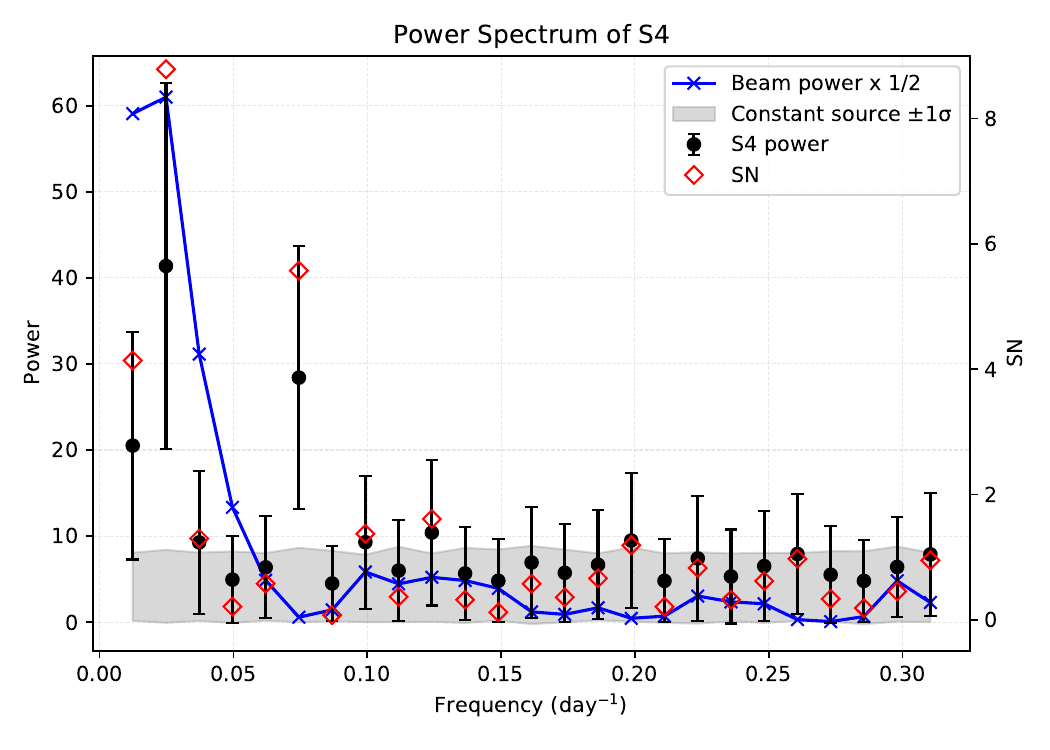}
\end{center}
\caption{Power spectrum of S3 (left) and S4 (right) as filled black circles, and their signal-to-noise (SN) as red open diamonds. The shaded region represents the range between $\pm 1$ standard deviation of a simulated source with constant flux and photometry errors that is same as S3 or S4. The spectrum of the beam area is shown in blue.}\label{powerspectrum}
\end{figure*}

Figure \ref{lightcurve} presents the S3 and S4 flux as a function of observation epoch (light curves), in Modified Julian Days (MJD).  The average flux is also shown as a dashed line. The 1$\sigma$ error bars are assigned as explained in subsection \ref{photometry}, but here the uncertainty coming from aperture photometry ($\sim 28\%$) dominate over errors from calibration. Individual images detect both S3 and S4 at a signal to noise over 5 for 49 EBs, and over 4 for the remaining 3 EBs.  We also show the variation of the beam size, which is a major source of systematic uncertainties when comparing individual images.  For both S3 and S4 we do not see any obvious dependence of the flux on beam size, which is consistent with both sources being (nearly) a point source.
The variations in the S3 flux seem to be mostly random, with no clear sign of variation on the timescale of several days.  For S4, the first $\sim 30$ days seem to show some coherent changes in flux. The latter 30 days show an apparent increase in flux towards the end of the run, but we caution that an increase is also observed in the beam size, which will be discussed later.

To further examine whether real time variations are present, and if there are characteristic timescales involved, we show in Figure \ref{powerspectrum} the power spectrum of the light curves.  The frequency resolution ($\sim 3$ days) was determined by the average of the time between epochs. Flux from individual EBs were normalized by their averages before performing Fourier transform.  The power is presented as the square of the amplitudes.  We estimated the uncertainties at each frequency bin using bootstrapping, where we changed the flux in each EB within their photometric uncertainties, assuming that the uncertainty at each EB is normally distributed.  Such realizations of the light curve and power spectrum were created 1000 times.  The values and error bars plotted in Figure \ref{powerspectrum} are the mean power and $1 \times$standard deviation of these realizations.

We also calculated the power spectrum for a simulated source with constant flux and noise that match the average flux and noise level of S3 and S4, observed at the same dates. The power spectrum of these simulated sources is shown as shaded regions in Figure \ref{powerspectrum}. The variation of power in the simulated source includes variations coming from both the flux uncertainty and the effect of uneven spacing of the observations over time.  The signal-to-noise (SN) of S3 power at each frequency is obtained by
\[
\rm{SN} = \frac{P_\mathrm{S3}-P_\mathrm{S3,sim}}{\sigma_\mathrm{S3,sim}},
\]
where $P_\mathrm{S3}$ is the power of S3, $P_\mathrm{S3,sim}$ is the power of a simulated source with constant flux, and $\sigma_\mathrm{S3,sim}$ is the $1 \times$standard deviation of the simulated source power, obtained using bootstrapping. The same calculation is done for S4 using the corresponding subscript.

We also show the power spectrum of the beam size fluctuations, scaled by $\times 0.5$ for clarity. While the flux of S3 and S4 should not be affected by the beam size due to its point-like nature, spatially correlated noise, which is intrinsic to interferometric measurements, can have systematic effects within the photometry aperture. As explained in section  \ref{photometry}, we chose an aperture of $0^{\prime \prime}.5$ to minimize this effect, but it may nevertheless be present.  Since the distribution of correlated noise (sidelobes) is strongly related to the beam size, we regard that flux correlated with the beam is a sign of systematics in the observation.

For S3, no notable peaks in the power spectrum are seen, except possibly at $0.3\ \mathrm{day^{-1}}$ with power of $17.6\pm 10.4$. The power deviates from the source with constant flux by $\rm{SN}=4.1$.
The frequency coincides with a small peak in the beam size power spectrum, that has power consistent with S3 within the error bars. It is also near the high frequency end of the analyzed range, corresponding to $\sim 3 $ days, the average cadence of the observations. Thus, we disregard this peak from further analysis.

For S4, there are peaks in the range $0.02 - 0.03\ \mathrm{day^{-1}}$.  Although they are the strongest peaks in the power spectrum, it coincides with a peak in the beam size power spectrum.  This component corresponds to 30-50 days, contributed largely by the gradual increase of flux and beam size in the latter half of the run.  It is likely that this peak is created by systematics, so we disregard peaks in the S4 power spectrum below $0.04\ \mathrm{day^{-1}}$ from further analyses.

We find a notable peak at $0.074\ \mathrm{day^{-1}}$ ($13\pm 3$ days) with power of $28.4\pm 15.4$.  The peak deviates from the source with constant flux by $\rm{SN}=5.6$, and cannot be explained by fluctuations in the beam size.  We regard this signal as a real sign of time variability.

Other than the peaks noted here, the power of the spectrum for both S3 and S4 maintain a relatively constant value within the power range expected for a source with constant flux.

\section{Discussion} \label{sec:discussion}
We find that the 3~mm continuum detections in the Antennae (overlap and southern NGC 4039) are mostly spatially extended and associated with star forming regions detected at other wavelengths.  Two sources, S4 associated with the stellar peak of NGC~4039, and S3 which is $\sim 100$ pc to the northeast of S4, stand out as compact sources.  S3 is slightly extended at a resolution of 10 pc, and is not found to be time variable.  S4 is unresolved even at 4 pc resolution.  It shows flux variability at a characteristic timescale of 13 days, which, if multiplied by speed of light, indicates that the phenomenon responsible for the variation has a spatial scale smaller than 0.01 pc. This size corresponds to $\sim10^4$ Schwarzschild radii for a $10^7M_\odot$ super-massive black-hole (SMBH).

\subsection{Multi-wavelength view of S3 and S4}\label{sec:multi}
The panels in Figure \ref{s3-s4} show the vicinity of S3 and S4 observed with JWST NIRCam and the $^{12}\mathrm{CO}(J=1-0)$ line included in the 114.6~GHz spw (Saito et al., in preparation).

The stellar continuum traced by JWST NIRCam $1.15\ \mathrm{\mu m}$ shows clearly that S4 is co-spatial with the stellar peak of NGC~4039.
JWST NIRCam F335M and F770W correspond to bands including PAH emission at $3.3~\mathrm{\mu m}$ and $7.7~\mathrm{\mu m}$, respectively.  Although caution must be taken that the underlying stellar and hot dust continuum have not been subtracted from these images, it is clear that while the $3.3~\mathrm{\mu m}$ roughly follows the stellar continuum, the $7.7~\mathrm{\mu m}$ is prominent around S3 but surprisingly weak around S4.

The overall abundance of PAH molecules is known to decrease in galaxies with AGN, both globally and within galaxies near the AGN \citep{Smith07, Sajina22}. This is broadly understood as a sign of PAH destruction by the strong UV and shocks, but the relative intensity of PAH features at different wavelengths varies depending on factors such as the PAH size, hardness of the radiation field, and the ionization parameter.  \cite{Brandl2009} show using Spitzer IRS spectra of Neon lines [NeIII] at $15.5\ \mathrm{\mu m}$ and [NeII] at $12.8\ \mathrm{\mu m}$, that the nucleus of NGC~4039 has a relatively hard radiation field with [NeIII]/[NeII]$\sim$0.5.  This ratio corresponds to a parameter space inhabited by AGN that have low $7.7~\mathrm{\mu m}$ to $11.3~\mathrm{\mu m}$ ratio \citep{Smith07}. A number of studies find that the $7.7~\mathrm{\mu m}$ can become weaker relative to $3.3~\mathrm{\mu m}$ when the PAH size is smaller or the radiation field is harder \citep{Rigopoulou21, Sidhu22, Chastenet2023, Zhang2025}. 

From the NIRCam images in Figure \ref{s3-s4}, the observed $7.7~\mathrm{\mu m}$/$3.3~\mathrm{\mu m}$ intensity ratio is $\sim 8$ and $\sim 0.5$ in S3 and S4, respectively.
Even if the stellar contribution to the $3.3~\mathrm{\mu m}$ in S4 is as large as 65\%, the largest value observed for a sample of galaxies \citep{Sandstrom23}, the $7.7~\mathrm{\mu m}$/$3.3~\mathrm{\mu m}$ ratio corrected for stellar contribution can only be elevated by a factor of 3 to $\sim 1.5$. Thus, S4 has a significantly lower $7.7~\mathrm{\mu m}$/$3.3~\mathrm{\mu m}$ ratio compared to S3.
Theoretical calculations by \cite{Rigopoulou21} show that being exposed to a harder radiation field (6 to 12~eV) can decrease $7.7~\mathrm{\mu m}$/$3.3~\mathrm{\mu m}$ from 0.6 to 0.4 for small PAH molecules, and from 11 to 2 for large PAH molecules. Thus, explaining the order of magnitude difference in $7.7~\mathrm{\mu m}$/$3.3~\mathrm{\mu m}$ between S3 and S4 likely requires not just a difference in the hardness of the radiation field, but also that PAH molecules in S4 are smaller than in S3, which could be due to destruction of large grains in S4. The relation of PAH line ratios to different physical states is still widely debated and thus we cannot draw definitive conclusions, but it is clear here that S3 and S4 are in very different environments.

 The lower panels in Figure \ref{s3-s4} show the integrated intensity (moment 0), the velocity field (moment 1), and the velocity dispersion (moment 2) image of the CO line. 
 A molecular arm-like structure in the vicinity of S3 has a blue-shifted line-of-sight velocity with respect to the component between the S3 and S4, and leads northwest-ward.  S4 does not have a directly overlapping molecular component, but a red-shifted arm-like structure leads southeast.  The integrated CO intensity of these structures typically exceed 1000 $\mathrm{K\ km/s}$, corresponding to a column density of $2\times 10^{23}\ \mathrm{cm^{-2}}$ or molecular surface mass density of $3\times 10^3\ \mathrm{M_\odot\ pc^{-2}}$ for a Galactic CO-to-$\mathrm{H_2}$ conversion factor of $2\times 10^{20}\ \mathrm{cm^{-2}(K\ km/s)^{-1}}$ \citep{strong96}. 
 The velocity dispersion within these structures is $20-30\ \mathrm{km\ s^{-1}}$, higher than its surroundings.

Association with an X-ray source would give strong constraints on the nature of S3 and S4.  Soft X-ray (0.1-10 keV) observations by Chandra \citep{Zezas2002a} 
detected a slightly extended X ray source (source \#29) in the southwest vicinity of the NGC~4039 nucleus.  The coordinate of this source with an updated astrometry \citep{Poutanen2013} is (R.A., Dec.)=(12h01m53.50s, -18d53m10.5s) with a positional uncertainty of $0^{\prime \prime}.15$, and shown in Figure \ref{s3-s4} along with its error circle.  The X-ray (0.1-10 keV) luminosity of this source is $L_X = 10^{39.57}\ \mathrm{erg\ s^{-1}}$ \citep{Poutanen2013}, making it one of the most luminous ultraluminous X-ray sources (ULXs) in the Antennae.

The most recent Chandra Source Catalog 2 (CRC2) \citep{CRC2} lists 2CXOJ120153.5-18 as being nearest to S4 at (R.A., Dec.)=(12h01m53.513s,-18d53m10.249s) with a positional uncertainty of $0^{\prime \prime}.29$, which encloses S4.  
This source has a steep spectrum with $\Gamma=2.5$, where $\Gamma$ is the photon index defined as $F_\mathrm{E}\propto E^{-\Gamma}$, and a best fit column density of $1.96\times 10^{21}\ \mathrm{cm^{-2}}$. Considering the large positional uncertainties, it is difficult to conclude whether this source in the CRC2 catalog is the same as \#29. Even if these two X ray sources are the same,
the unusually soft X-ray spectrum and the high millimeter luminosity of S4 ($\sim10^{37}~\mathrm{erg~s^{-1}}$: see section \ref{origin}) make a physical association of the X-ray source with S4 non-trivial.  


\subsection{The Origin of millimeter emission}\label{origin}
Millimeter continuum emission can be observed as thermal (free-free bremsstrahlung or dust) or non-thermal (synchrotron) emission. 
Thermal free-free emission typically has a shallow spectral slope ($\sim -0.1$), and is detected in star-forming regions, particularly from young massive stellar clusters (YMCs).  Emission from YMCs is not expected to vary with time, at least at the timescales probed here (3 to 30 days).  Therefore, free-free emission is unlikely as the origin of millimeter emission from S4, but still a viable mechanism for S3.

The millimeter emission is unlikely to be thermal dust emission.  For a typical Milky Way dust emissivity of $0.37\ \mathrm{m^2\ kg^{-1}}$ at $250\ \mathrm{\mu m}$ \citep{Bianchi19}, dust emissivity index of 2, and a dust temperature of 30 K, the observed flux of  $\sim 150$ $\mathrm{\mu Jy}$ at 100 GHz corresponds to a dust mass of $\sim 10^4\ \mathrm{M_\odot}$ (i.e., molecular cloud mass of $\sim 10^6\ \mathrm{M_\odot}$). This is two orders of magnitude more massive than would be expected from a typical 10 parsec diameter molecular cloud following the size-mass relation.
The spectral index of S3 and S4 (see section \ref{sec:index}) is flat within the errors, which also suggests negligible contribution from dust. Contribution of Anomalous Microwave Emission (AME) from, e.g., spinning dust \citep{Planck11} may alter the spectral index, but emission at $~100$ GHz is still expected to be dominated by normal thermal dust with positive spectral index, so AME is unlikely to explain emission from these sources.

Non-thermal emission is produced by energetic objects such as X-ray binaries, supernova remnants (SNR), and AGN.  The spectral slope of synchrotron emission is generally steeper than thermal emission, ranging from $-0.5$ to $-1.5$, depending on the electron population.
The 3~mm flux given in Table \ref{sources} corresponds to 100~GHz luminosities ($\nu L_\mathrm{\nu}$) of $8.8\times 10^{36}\ \mathrm{erg\ s^{-1}}$ and $9.1\times 10^{36}\ \mathrm{erg\ s^{-1}}$ for S3 and S4, respectively.  
This is an order of magnitude more luminous than a typical luminous SNR at 1.4~GHz \citep{Chomiuk09, Cseh2012} which is about $\sim 10^{35}\ \mathrm{erg\ s^{-1}}$, if we assume a reasonable spectral index of $\alpha = -0.6$ to convert from 1.4~GHz to 100~GHz.  SNRs are therefore unlikely as the origin of millimeter emission in S3 or S4.

Could S3 and/or S4 be X-ray binary systems with synchrotron jets (i.e., a microquasar)?
The slight southwest-northeast extension of millimeter emission in S3 is indicative of such a small scale jet. The characteristic 13 day variation in S4 seems consistent with that expected for binary systems, which would tempt us to believe that the nearby Chandra source 2CXOJ120153.5-185310 could be associated with S4.  We find, however, that
the observed millimeter luminosities of S3 and S4 are inconsistent with microquasars.  ALMA observations of the Galactic microquasar SS433 \citep{Marti2018} at 230~GHz measure a flux of 70-80 mJy, with rapid time variation within minutes.  Assuming a distance of 5.5 kpc \citep{Lockman2007} and a flat spectrum, the corresponding 100~GHz luminosity is $2.5-3.3 \times 10^{32}\ \mathrm{erg\ s^{-1}}$.  Similarly, GRS1915+105 was observed at 100~GHz and measured 2.5-2.9 mJy \citep{Koljonen2021}.  Assuming a distance of 11 kpc \citep{Fender1999}, the 100~GHz luminosity is $4.2 \times 10^{31}\ \mathrm{erg\ s^{-1}}$.  Other millimeter detection of X ray binaries give luminosities ranging $10^{27-32}\ \mathrm{erg\ s^{-1}}$ \citep{Gallo19, DiazTrigo21}.
The luminosities of S3 and S4 ($\sim 10^{37}\ \mathrm{erg\ s^{-1}}$) are orders of magnitude higher than typical microquasars; therefore, we cannot attribute the millimeter flux of S3 or S4 to X-ray binary systems.


We finally assess whether S3 and S4 are consistent with an AGN.
 The characteristic timescale of flux variation in S4 (13 days) indicates that the mechanism responsible for the emission is smaller than 0.01 pc.  The flux of S4 in the uniform image observed at $0^{\prime \prime}.04$ (4 pc) resolution ($117~ \mathrm{\mu Jy}$ at 100 GHz), corresponds to a brightness temperature of $8.2$ K.  If we assume that this emission comes entirely from a source with size smaller than 0.01 pc, we obtain a lower limit for the brightness temperature of $1.3\times 10^6\ \mathrm{K}$, which strongly favors a non-thermal origin for S4. Coronal synchrotron emission from the vicinity of an SMBH has been theoretically investigated \citep{Inoue2014}, and later confirmed by ALMA observations (\citealp{Inoue2018, Michiyama2024, Shablovinskaya2024}, but see also \citealp{Yamada24, Hankla25}). The SED typically exhibits a peak at frequencies of several tens of GHz, which likely corresponds to the synchrotron self-absorption frequency.  The SED is dominated by the synchrotron jet at lower frequencies, so the centimeter-millimeter SED is complex.  The shallow cm-mm spectral slope of S3 and S4, with their large uncertainties, does not allow for any conclusive discussion on the shape of the SED, and more detailed observations over a wider frequency range is necessary to constrain the emission mechanism.

If S3 or S4 are AGN, we would expect \textit{hard} X ray to be detected. A correlation is known to exist between millimeter continuum (100 GHz) emission and hard X ray (14-195 keV) emission in radio-quiet AGNs \citep{Kawamuro2022, Ricci2023}.  The millimeter luminosity of S3 and S4 ($\sim 10^{37}\ \mathrm{erg\ s^{-1}}$) corresponds to the lowest observed range in \cite{Ricci2023}, where much of the Seyfert galaxies reside. If we assume the same correlation for S3 and S4, we estimate the X ray luminosity to be $\sim 10^{42.1}\ \mathrm{erg\ s^{-1}}$. This is inconsistent with the non-detection in the SWIFT/BAT catalog \citep{Oh2018}, where its sensitivity of $8.4\times 10^{-12}\ \mathrm{erg\ s^{-1}\ cm^{-2}}$ over 90\% of the sky corresponds to a sensitivity of $\sim 5\times 10^{41}\ \mathrm{erg\ s^{-1}}$.

Assuming that S3 and S4 follow the mm-X ray relation, the non-detection in the X ray can be explained if the AGN were Compton thick, i.e., with such a high hydrogen column density that could lower the X ray luminosity by an order of magnitude. \cite{Brandl2009} detected warm $\mathrm{H_2}$ at the nucleus of NGC~4039 using Spitzer IRS spectra, and estimated a warm molecular gas mass of $3.56 \times 10^6\ \mathrm{M_\odot}$ within a slit size corresponding to $0.5\times 1.2\ \mathrm{kpc^2}$.  The total ISM mass would be much larger if we include cold gas. If more than $\sim 10\%$ of this warm molecular gas is confined within the line of sight towards S4, the column density $N(\mathrm{H})$ would become larger than $10^{24}\ \mathrm{cm^{-2}}$, making S4 Compton-thick.  It is an open question whether the AGN candidates in this study should follow the mm-X ray relation, however. S3 and S4 being Compton-thick remains only a possibility.

\section{Conclusion}
We used ALMA observations at 100 GHz, executed 52 times over a period of 2.5 months, to search for time variability of millimeter emission.  

Two compact sources S3 and S4, are found near the nuclei of the southern galaxy NGC\ 4039. 
S3 is slightly extended at 10 parsec resolution.
Its luminosity is at least an order of magnitude higher than typical bright SNRs , X-ray binaries, or dust associated with molecular clouds.  It does not show any signs of time variability. Thermal free-free emission from a YMC is a viable origin of S3, although the possibility of an AGN is not rejected.

S4 is a compact source associated with the stellar peak of NGC~4039, and remains unresolved at 4~pc resolution.  It is time variable and has a characteristic timescale of 13 days, which indicates that the structure giving rise to the millimeter flux has a size of $< 0.01$ parsecs.  Assuming this size, the brightness temperature at 100~GHz is $>10^6\ \mathrm{K}$, suggesting non-thermal origin.  The time variability excludes dust, YMC, SNR or other star formation origin for S4. Its luminosity is also inconsistent with dust, SNR or X-ray binaries. These observations indicate that S4 is a strong candidate for an AGN.
The $7.7~\mathrm{\mu m}$ to $3.3~\mathrm{\mu m}$ ratio hints at a hard radiation field and possible destruction of PAH molecules. Molecular gas traced by CO show arm-like structures, but CO is not directly associated with S4.

The flux of S3 and S4 between different spectral windows in the ALMA observations
are consistent with flat or shallow slope, but the large flux uncertainties do not allow for any robust conclusions regarding the SED.
Neither S3 nor S4 are detected in hard X-ray, which could be explained if the AGN is a Compton-thick AGN.  Our observations could indicate that there could be an obscured AGN already in an early-phase interacting galaxy.

Observations over a wider frequency range can constrain the SED of S3 and S4. Follow-up observations of IR spectroscopy with JWST, and sensitive X ray telescopes like NuSTAR may be able to further reveal the nature of these sources.





\begin{acknowledgments}
S.K. is supported by JSPS KAKENHI grant No. JP25K07371.
T.S. is supported by Daiichi-Sankyo ``Habataku" Support Program for the Next Generation of Researchers.
T.M. is supported by JSPS KAKENHI grant No. JP25K17441.
Y.I. is supported by JSPS KAKENHI grant No. JP22K18277 and JP26H00604.
F.M. is supported by JSPS KAKENHI grant No. JP23K13142 and JP23K20035.
This paper makes use of the following ALMA data: ADS/JAO.ALMA\#2022.A.00032.S.
ALMA is a partnership of ESO (representing its member states), NSF (USA) and NINS (Japan), together with NRC (Canada), NSTC and ASIAA (Taiwan), and KASI (Republic of Korea), in cooperation with the Republic of Chile. The Joint ALMA Observatory is operated by ESO, AUI/NRAO and NAOJ. Data analysis was in part carried out on the common use data analysis computer system at the Astronomy Data Center, ADC, of the National Astronomical Observatory of Japan. This work is based in part on observations made with the NASA/ESA/CSA James Webb Space Telescope. The data were obtained from the Mikulski Archive for Space Telescopes at the Space Telescope Science Institute, which is operated by the Association of Universities for Research in Astronomy, Inc., under NASA contract NAS 5-03127 for JWST. These observations are associated with program \#1995 and \#2581.
\end{acknowledgments}

\vspace{5mm}
\facilities{ALMA, VLA}

\software{
ALMA Interferometric Pipeline \citep{Hunter23},
CASA \citep{CASA22},
}

\appendix
\section{list of individual Execution Blocks}

\startlongtable
\begin{deluxetable*}{lclccccccrc}
\tablecaption{List of QA0-passed data obtained for this project\label{data}}
\tablewidth{0pt}
\tablehead{
\colhead{Date \& Time} & \colhead{$t_{\rm ON}$} & \colhead{Execution Block} & \colhead{$N_{\rm ant}$} & \colhead{$L_{\rm baseline}$} & \colhead{$\theta_\mathrm{MRS}$}& \colhead{PWV} & \colhead{Calibrator} & \multicolumn{2}{c}{beam} & \colhead{noise} \\
\colhead{2023 UT} & \colhead{min.} & \colhead{uid://A002/} & & \colhead{m} & \colhead{$\prime \prime$} & \colhead{mm} & \colhead{Bandpass} &  \colhead{$\prime \prime \times \prime \prime$} &  \colhead{P.A.} &  \colhead{$\mu$Jy/b}
}
\startlongtable
\startdata
Sep 05 16:54 & 51.4 & X10c5c57/X47b2  & 43 & 83--12271 & 1.09 & 0.62 & J1256$-$0547 & 0.085$\times$0.076 & 50.4    & 16.4 \\
Sep 07 16:35 & 51.5 & X10c6fac/X65c6  & 48 & 83--14850 & 1.30 & 0.90 & J1256$-$0547 & 0.095$\times$0.081 & 53.9    & 14.9 \\
Sep 08 20:15 & 51.5 & X10c6fac/Xb347  & 44 & 83--12271 & 1.06 & 1.26 & J1256$-$0547 & 0.091$\times$0.083 & $-$83.1 & 16.8 \\
Sep 08 22:09 & 51.5 & X10c6fac/Xb92b  & 44 & 83--12271 & 1.20 & 1.42 & J1256$-$0547 & 0.14$\times$0.083 & $-$77.1  & 18.9 \\
Sep 09 16:39 & 51.5 &  X10c9905/X408   & 45 & 83--12271 & 1.34 & 0.80 & J1256$-$0547 & 0.097$\times$0.094 & $-$76.9 & 15.6 \\
Sep 09 19:23 & 51.4 &  X10c9905/Xb46   & 45 & 83--12271 & 1.22 & 0.91 & J1256$-$0547 & 0.094$\times$0.087 & $-$33.1 & 15.6 \\
Sep 13 16:01 & 36.6 &  X10ca5ad/Xd944  & 45 & 66--11886 & 1.51 & 0.85 & J1256$-$0547 & 0.14$\times$0.11  & 70.5    & 17.2 \\
Sep 14 15:49 & 51.5 &  X10cc13c/X67d1  & 44 & 83--9743  & 1.57 & 1.24 & J1256$-$0547 & 0.16$\times$0.12  & 86.3    & 17.3 \\
Sep 14 17:46 & 51.4 &  X10cc13c/X79a3  & 44 & 83--9743  & 1.43 & 1.46 & J1256$-$0547 & 0.13$\times$0.12  & $-$80.8 & 16.8 \\
Sep 14 19:35 & 51.5 &  X10cc13c/X855a  & 44 & 83--9743  & 1.44 & 1.43 & J1256$-$0547 & 0.14$\times$0.12  & $-$66.4 & 17.1 \\
Sep 18 14:34 & 51.4 &  X10ceb22/X67cb  & 44 & 89--8547  & 2.31 & 1.35 & J1037$-$2934 & 0.24$\times$0.16  & 84.2    & 18.3 \\
Sep 18 16:32 & 51.4 &  X10ceb22/X702e  & 43 & 91--8547  & 1.89 & 1.17 & J1256$-$0547 & 0.17$\times$0.15  & $-$67.4 & 18.0 \\
Sep 18 17:56 & 37.3 &  X10ceb22/X77e4  & 43 & 91--8547  & 1.77 & 1.02 & J1037$-$2934 & 0.17$\times$0.14  & $-$51.9 & 17.6 \\
Sep 19 15:37 & 51.5 &  X10ceb22/Xe75d  & 41 & 89--8282  & 2.03 & 2.01 & J1037$-$2934 & 0.19$\times$0.14  & $-$86.0 & 19.7 \\
Sep 20 15:26 & 51.5 &  X10ceb22/X15cf6 & 44 & 91--8547  & 1.88 & 2.04 & J1037$-$2934 & 0.18$\times$0.14  & 76.0    & 18.6 \\
Sep 21 18:48 & 51.5 &  X10d12a2/X3ae9  & 41 & 97--8547  & 1.74 & 1.02 & J1037$-$2934 & 0.15$\times$0.14  & 89.0    & 17.9 \\
Sep 22 18:40 & 51.5 &  X10d12a2/Xb6c2  & 43 & 97--8547  & 1.82 & 0.91 & J1037$-$2934 & 0.16$\times$0.14  & 82.0    & 17.4 \\
Sep 23 15:59 & 8.9 &  X10d12a2/X11779  & 42 & 89--8282  & 1.98 & 0.33 & J1037$-$2934 & 0.19$\times$0.15  & $-$76.9 & 35.1 \\
Sep 25 18:56 & 51.5 &  X10d4e2e/X2ae   & 43 & 91--8547  & 1.87 & 3.24 & J1037$-$2934 & 0.18$\times$0.14  & $-$53.9 & 20.4 \\
Sep 25 20:42 & 51.5 &  X10d4e2e/Xb98   & 43 & 91--8547  & 2.12 & 3.25 & J1107$-$4449 & 0.25$\times$0.15  & $-$61.7 & 22.2 \\
Sep 26 15:42 & 51.5 &  X10d4e2e/X8322  & 42 & 91--8547  & 1.97 & 1.68 & J1037$-$2934 & 0.18$\times$0.15  & 82.8    & 19.9 \\
Sep 26 17:41 & 51.5 &  X10d4e2e/X8bf1  & 42 & 91--8547  & 1.84 & 1.76 & J1037$-$2934 & 0.16$\times$0.14  & $-$83.0 & 19.5 \\
Sep 28 17:01 & 51.5 &  X10d4e2e/X15558 & 47 & 91--8547  & 1.86 & 0.48 & J1037$-$2934 & 0.18$\times$0.14  & 71.3    & 15.7 \\
Sep 29 15:16 & 51.5 &  X10d4e2e/X1c482 & 43 & 89--8547  & 1.97 & 0.52 & J1037$-$2934 & 0.19$\times$0.14  & 59.5    & 17.5 \\
Sep 29 17:07 & 51.6 &  X10d4e2e/X1c9ef & 43 & 89--8547  & 1.83 & 0.98 & J1037$-$2934 & 0.18$\times$0.14  & 61.4    & 17.4 \\
Sep 29 20:08 & 51.5 &  X10d4e2e/X1d230 & 45 & 91--8547  & 2.04 & 1.26 & J1037$-$2934 & 0.22$\times$0.14  & 89.9    & 20.0 \\
Sep 30 15:31 & 51.5 &  X10d4e2e/X241f6 & 44 & 91--8547  & 1.96 & 1.47 & J1037$-$2934 & 0.21$\times$0.16  & $-$83.6 & 17.6 \\
Oct 12 18:19 & 51.5 &  X10e09b4/X82d1  & 41 & 91--8547  & 1.99 & 0.83 & J1037$-$2934 & 0.20$\times$0.15  & $-$89.9 & 18.2 \\
Oct 14 14:06 & 7.4 &  X10e2702/X74b1   & 39 & 97--8282  & 1.96 & 1.08 & J1037$-$2934 & 0.19$\times$0.14  & 64.1    & 42.6 \\
Oct 16 16:07 & 51.4 &  X10e492f/X121   & 42 & 91--8547  & 1.76 & 3.86 & J1037$-$2934 & 0.16$\times$0.14  & 85.0    & 21.0 \\
Oct 16 19:11 & 51.5 &  X10e492f/X1262  & 42 & 91--8547  & 1.94 & 3.98 & J1037$-$2934 & 0.22$\times$0.13  & $-$79.0 & 24.1 \\
Oct 17 14:05 & 47.3 &  X10e492f/X7ce3  & 43 & 91--8547  & 1.87 & 4.33 & J1037$-$2934 & 0.17$\times$0.13  & 65.1    & 20.0 \\
Oct 17 19:35 & 51.5 &  X10e492f/X9a95  & 42 & 91--8547  & 2.24 & 4.00 & J1107$-$4449 & 0.25$\times$0.14  & $-$77.3 & 24.3 \\
Oct 20 15:59 & 51.5 &  X10e6d25/X6d26  & 42 & 91--8282  & 1.91 & 2.38 & J1229+0203   & 0.18$\times$0.14  & 73.8    & 19.8 \\
Oct 20 17:59 & 51.4 &  X10e6d25/X7b2d  & 43 & 91--8282  & 1.93 & 2.56 & J1229+0203   & 0.21$\times$0.14  & $-$86.0 & 20.3 \\
Oct 24 16:48 & 51.5 &  X10e9b60/X7c3c  & 42 & 91--8547  & 1.72 & 0.95 & J1229+0203   & 0.15$\times$0.14  & $-$56.2 & 19.4 \\
Oct 25 12:57 & 51.5 &  X10e9b60/Xf642  & 44 & 113--8282 & 2.05 & 0.83 & J1229+0203   & 0.21$\times$0.16  & 89.7    & 18.6 \\
Oct 26 14:49 & 51.4 &  X10eb7e9/X53e2  & 43 & 91--8282  & 1.87 & 0.53 & J1229+0203   & 0.19$\times$0.15  & $-$89.0 & 18.4 \\
Oct 26 16:55 & 51.5 &  X10eb7e9/X5b5c  & 43 & 91--8282  & 1.78 & 0.59 & J1229+0203   & 0.27$\times$0.14  & $-$47.2 & 21.6 \\
Oct 28 14:22 & 29.7 &  X10eb7e9/Xc44f  & 42 & 91--8282  & 1.87& 0.37 & J1229+0203   & 0.20$\times$0.14  & 67.8    & 20.5 \\
Oct 31 17:38 & 51.5 &  X10ed869/Xaccf  & 45 & 66--8282  & 2.63 & 0.65 & J1256$-$0547 & 0.27$\times$0.20  & $-$77.7 & 17.9 \\
Nov 01 16:33 & 47.2 &  X10ed869/X14e1a & 45 & 85--8282  & 2.21 & 0.71 & J1256$-$0547 & 0.23$\times$0.17  & 76.4    & 17.5 \\
Nov 06 13:57 & 51.5 &  X10f3768/X3fc   & 41 & 85--6582  & 2.67 & 4.91 & J1256$-$0547 & 0.22$\times$0.19  & 80.7    & 21.7 \\
Nov 06 15:15 & 29.9 &  X10f3768/Xef3   & 41 & 85--6582  & 2.63 & 4.80 & J1256$-$0547 & 0.23$\times$0.19  & $-$80.6 & 24.3 \\
Nov 06 17:27 & 51.5 &  X10f3768/X15c8  & 41 & 45--6582  & 2.89 & 4.53 & J1256$-$0547 & 0.28$\times$0.19  & $-$77.8 & 23.9 \\
Nov 07 18:08 & 51.5 &  X10f3768/X92ce  & 46 & 30--6582  & 3.29 & 5.16 & J1256$-$0547 & 0.39$\times$0.22  & $-$73.6 & 25.2 \\
Nov 08 11:21 & 12.1 &  X10f3768/Xc9fa  & 49 & 30--5185  & 3.35 & 4.56 & J1256$-$0547 & 0.30$\times$0.22  & 72.0    & 31.6 \\
Nov 08 12:32 & 30.9 &  X10f3768/Xcc54  & 49 & 30--5185  & 3.17 & 4.87 & J1256$-$0547 & 0.27$\times$0.21  & 71.9    & 20.4 \\
Nov 08 14:38 & 51.5 &  X10f3768/Xd1c1  & 46 & 45--5185  & 2.96 & 5.22 & J1256$-$0547 & 0.29$\times$0.22  & 76.8    & 21.0 \\
Nov 09 12:06 & 51.5 &  X10f3768/X119de & 51 & 30--5185  & 3.33 & 5.45 & J1256$-$0547 & 0.31$\times$0.22  & 71.7    & 20.6 \\
Nov 09 16:02 & 51.4 &  X10f3768/X12daf & 48 & 30--5185  & 2.85 & 4.72 & J1256$-$0547 & 0.32$\times$0.22  & 79.1    & 20.6 \\
Nov 23 15:49 & 20.9 &  X10fdea7/X39d2  & 42 & 45--5185  & 3.19 & 2.72 & J1256$-$0547 & 0.37$\times$0.20  & $-$75.3 & 28.6 \\
\enddata
\tablecomments{
Column 1: Observation date. Column 2: Time on target.  Column 3: Execution Block UID.  Column 4: Number of antennas.  Column 5: Length of the shortest to longest baselines.  Column 6: The Maximum Recoverable Scale, calculated following equation 7.6 in the ALMA Cycle 9 Technical Handbook \citep{ALMA-TH}. Column 7: Precipitable Water Vapor of the observations. Column 8: Name of Bandpass calibrator.  Column 9: Major and minor axis FWHM, and position angle of the synthesized beam. Column 10: r.m.s. noise level.
}
\end{deluxetable*}

\bibliography{sample631}{}

@ARTICLE{Leroy21,
       author = {{Leroy}, Adam K. and {Hughes}, Annie and {Liu}, Daizhong and {Pety}, J{\'e}r{\^o}me and {Rosolowsky}, Erik and {Saito}, Toshiki and {Schinnerer}, Eva and {Schruba}, Andreas and {Usero}, Antonio and {Faesi}, Christopher M. and {Herrera}, Cinthya N. and {Chevance}, M{\'e}lanie and {Hygate}, Alexander P.~S. and {Kepley}, Amanda A. and {Koch}, Eric W. and {Querejeta}, Miguel and {Sliwa}, Kazimierz and {Will}, David and {Wilson}, Christine D. and {Anand}, Gagandeep S. and {Barnes}, Ashley and {Belfiore}, Francesco and {Be{\v{s}}li{\'c}}, Ivana and {Bigiel}, Frank and {Blanc}, Guillermo A. and {Bolatto}, Alberto D. and {Boquien}, M{\`e}d{\`e}ric and {Cao}, Yixian and {Chandar}, Rupali and {Chastenet}, J{\'e}r{\'e}my and {Chiang}, I-Da and {Congiu}, Enrico and {Dale}, Daniel A. and {Deger}, Sinan and {den Brok}, Jakob S. and {Eibensteiner}, Cosima and {Emsellem}, Eric and {Garc{\'\i}a-Rodr{\'\i}guez}, Axel and {Glover}, Simon C.~O. and {Grasha}, Kathryn and {Groves}, Brent and {Henshaw}, Jonathan D. and {Jim{\'e}nez Donaire}, Mar{\'\i}a J. and {Kim}, Jaeyeon and {Klessen}, Ralf S. and {Kreckel}, Kathryn and {Kruijssen}, J.~M. Diederik and {Larson}, Kirsten L. and {Lee}, Janice C. and {Mayker}, Ness and {McElroy}, Rebecca and {Meidt}, Sharon E. and {Mok}, Angus and {Pan}, Hsi-An and {Puschnig}, Johannes and {Razza}, Alessandro and {S{\'a}nchez-Bl'azquez}, Patricia and {Sandstrom}, Karin M. and {Santoro}, Francesco and {Sardone}, Amy and {Scheuermann}, Fabian and {Sun}, Jiayi and {Thilker}, David A. and {Turner}, Jordan A. and {Ubeda}, Leonardo and {Utomo}, Dyas and {Watkins}, Elizabeth J. and {Williams}, Thomas G.},
        title = "{PHANGS-ALMA Data Processing and Pipeline}",
      journal = {\apjs},
         year = 2021,
        month = jul,
       volume = {255},
       number = {1},
          eid = {19},
        pages = {19},
          doi = {10.3847/1538-4365/abec80},
archivePrefix = {arXiv},
       eprint = {2104.07665},
 primaryClass = {astro-ph.IM},
       adsurl = {https://ui.adsabs.harvard.edu/abs/2021ApJS..255...19L}
}

@ARTICLE{Bianchi19,
       author = {{Bianchi}, S. and {Casasola}, V. and {Baes}, M. and {Clark}, C.~J.~R. and {Corbelli}, E. and {Davies}, J.~I. and {De Looze}, I. and {De Vis}, P. and {Dobbels}, W. and {Galametz}, M. and {Galliano}, F. and {Jones}, A.~P. and {Madden}, S.~C. and {Magrini}, L. and {Mosenkov}, A. and {Nersesian}, A. and {Viaene}, S. and {Xilouris}, E.~M. and {Ysard}, N.},
        title = "{Dust emissivity and absorption cross section in DustPedia late-type galaxies}",
      journal = {\aap},
         year = 2019,
        month = nov,
       volume = {631},
          eid = {A102},
        pages = {A102},
          doi = {10.1051/0004-6361/201936314},
archivePrefix = {arXiv},
       eprint = {1909.12692},
 primaryClass = {astro-ph.GA},
       adsurl = {https://ui.adsabs.harvard.edu/abs/2019A&A...631A.102B}
}

@ARTICLE{Strong96,
       author = {{Strong}, A.~W. and {Mattox}, J.~R.},
        title = "{Gradient model analysis of EGRET diffuse Galactic {\ensuremath{\gamma}}-ray emission.}",
      journal = {\aap},
         year = 1996,
        month = apr,
       volume = {308},
        pages = {L21-L24},
       adsurl = {https://ui.adsabs.harvard.edu/abs/1996A&A...308L..21S}
}

@ARTICLE{Planck11,
       author = {{Planck Collaboration} and {Ade}, P.~A.~R. and {Aghanim}, N. and {Arnaud}, M. and {Ashdown}, M. and {Aumont}, J. and {Baccigalupi}, C. and {Balbi}, A. and {Banday}, A.~J. and {Barreiro}, R.~B. and {Bartlett}, J.~G. and {Battaner}, E. and {Benabed}, K. and {Beno{\^\i}t}, A. and {Bernard}, J.-P. and {Bersanelli}, M. and {Bhatia}, R. and {Bock}, J.~J. and {Bonaldi}, A. and {Bond}, J.~R. and {Borrill}, J. and {Bouchet}, F.~R. and {Boulanger}, F. and {Bucher}, M. and {Burigana}, C. and {Cabella}, P. and {Cappellini}, B. and {Cardoso}, J.-F. and {Casassus}, S. and {Catalano}, A. and {Cay{\'o}n}, L. and {Challinor}, A. and {Chamballu}, A. and {Chary}, R.-R. and {Chen}, X. and {Chiang}, L.-Y. and {Chiang}, C. and {Christensen}, P.~R. and {Clements}, D.~L. and {Colombi}, S. and {Couchot}, F. and {Coulais}, A. and {Crill}, B.~P. and {Cuttaia}, F. and {Danese}, L. and {Davies}, R.~D. and {Davis}, R.~J. and {de Bernardis}, P. and {de Gasperis}, G. and {de Rosa}, A. and {de Zotti}, G. and {Delabrouille}, J. and {Delouis}, J.-M. and {Dickinson}, C. and {Donzelli}, S. and {Dor{\'e}}, O. and {D{\"o}rl}, U. and {Douspis}, M. and {Dupac}, X. and {Efstathiou}, G. and {En{\ss}lin}, T.~A. and {Eriksen}, H.~K. and {Finelli}, F. and {Forni}, O. and {Frailis}, M. and {Franceschi}, E. and {Galeotta}, S. and {Ganga}, K. and {G{\'e}nova-Santos}, R.~T. and {Giard}, M. and {Giardino}, G. and {Giraud-H{\'e}raud}, Y. and {Gonz{\'a}lez-Nuevo}, J. and {G{\'o}rski}, K.~M. and {Gratton}, S. and {Gregorio}, A. and {Gruppuso}, A. and {Hansen}, F.~K. and {Harrison}, D. and {Helou}, G. and {Henrot-Versill{\'e}}, S. and {Herranz}, D. and {Hildebrandt}, S.~R. and {Hivon}, E. and {Hobson}, M. and {Holmes}, W.~A. and {Hovest}, W. and {Hoyland}, R.~J. and {Huffenberger}, K.~M. and {Jaffe}, T.~R. and {Jaffe}, A.~H. and {Jones}, W.~C. and {Juvela}, M. and {Keih{\"a}nen}, E. and {Keskitalo}, R. and {Kisner}, T.~S. and {Kneissl}, R. and {Knox}, L. and {Kurki-Suonio}, H. and {Lagache}, G. and {L{\"a}hteenm{\"a}ki}, A. and {Lamarre}, J.-M. and {Lasenby}, A. and {Laureijs}, R.~J. and {Lawrence}, C.~R. and {Leach}, S. and {Leonardi}, R. and {Lilje}, P.~B. and {Linden-V{\o}rnle}, M. and {L{\'o}pez-Caniego}, M. and {Lubin}, P.~M. and {Mac{\'\i}as-P{\'e}rez}, J.~F. and {MacTavish}, C.~J. and {Maffei}, B. and {Maino}, D. and {Mandolesi}, N. and {Mann}, R. and {Maris}, M. and {Marshall}, D.~J. and {Mart{\'\i}nez-Gonz{\'a}lez}, E. and {Masi}, S. and {Matarrese}, S. and {Matthai}, F. and {Mazzotta}, P. and {McGehee}, P. and {Meinhold}, P.~R. and {Melchiorri}, A. and {Mendes}, L. and {Mennella}, A. and {Mitra}, S. and {Miville-Desch{\^e}nes}, M.-A. and {Moneti}, A. and {Montier}, L. and {Morgante}, G. and {Mortlock}, D. and {Munshi}, D. and {Murphy}, A. and {Naselsky}, P. and {Natoli}, P. and {Netterfield}, C.~B. and {N{\o}rgaard-Nielsen}, H.~U. and {Noviello}, F. and {Novikov}, D. and {Novikov}, I. and {O'Dwyer}, I.~J. and {Osborne}, S. and {Pajot}, F. and {Paladini}, R. and {Partridge}, B. and {Pasian}, F. and {Patanchon}, G. and {Pearson}, T.~J. and {Peel}, M. and {Perdereau}, O. and {Perotto}, L. and {Perrotta}, F. and {Piacentini}, F. and {Piat}, M. and {Plaszczynski}, S. and {Platania}, P. and {Pointecouteau}, E. and {Polenta}, G. and {Ponthieu}, N. and {Poutanen}, T. and {Pr{\'e}zeau}, G. and {Procopio}, P. and {Prunet}, S. and {Puget}, J.-L. and {Reach}, W.~T. and {Rebolo}, R. and {Reich}, W. and {Reinecke}, M. and {Renault}, C. and {Ricciardi}, S. and {Riller}, T. and {Ristorcelli}, I. and {Rocha}, G. and {Rosset}, C. and {Rowan-Robinson}, M. and {Rubi{\~n}o-Mart{\'\i}n}, J.~A. and {Rusholme}, B. and {Sandri}, M. and {Santos}, D. and {Savini}, G. and {Scott}, D. and {Seiffert}, M.~D. and {Shellard}, P. and {Smoot}, G.~F. and {Starck}, J.-L. and {Stivoli}, F. and {Stolyarov}, V. and {Stompor}, R. and {Sudiwala}, R. and {Sygnet}, J.-F. and {Tauber}, J.~A. and {Terenzi}, L. and {Toffolatti}, L. and {Tomasi}, M. and {Torre}, J.-P. and {Tristram}, M.},
        title = "{Planck early results. XX. New light on anomalous microwave emission from spinning dust grains}",
      journal = {\aap},
         year = 2011,
        month = dec,
       volume = {536},
          eid = {A20},
        pages = {A20},
          doi = {10.1051/0004-6361/201116470},
archivePrefix = {arXiv},
       eprint = {1101.2031},
 primaryClass = {astro-ph.GA},
       adsurl = {https://ui.adsabs.harvard.edu/abs/2011A&A...536A..20P}
}

@ARTICLE{Farren21,
       author = {{Farren}, Gerrit S. and {Partridge}, Bruce and {Kneissl}, R{\"u}diger and {Aiola}, Simone and {Datta}, Rahul and {Gralla}, Megan and {Li}, Yaqiong},
        title = "{Confirming the Calibration of ALMA Using Planck Observations}",
      journal = {\apjs},
         year = 2021,
        month = sep,
       volume = {256},
       number = {1},
          eid = {19},
        pages = {19},
          doi = {10.3847/1538-4365/ac090d},
archivePrefix = {arXiv},
       eprint = {2102.05079},
 primaryClass = {astro-ph.IM},
       adsurl = {https://ui.adsabs.harvard.edu/abs/2021ApJS..256...19F}
}

@ARTICLE{ALMA-TH,
       author = {{Cortes}, P.C. and {Remijan}, A.  and {Hales}, A.  and {Carpenter}, J.  and {Dent}, W.  and {Kameno}, S.  and {Loomis}, R.  and {Vila-Vilaro}, B.  and {Bigg}, A.  and {Miotello}, A.  and {Vlahakis}, C.  and {Rosen}, R.  and {Stoehr}, F.  and {Saini}, K.},
        title = "{ALMA Technical Handbook}",
      journal = {ALMA Doc. 9.3, ver. 1.0},
         year = 2022,
         isbn = {978-3-923524-66-2}
}

@ARTICLE{Guzman19,
       author = {{Guzm{\'a}n}, A.~E. and {Verdugo}, C. and {Nagai}, H. and {Contreras}, Y. and {Marinello}, G. and {Kneissl}, R. and {Nakanishi}, K. and {Ueda}, J.},
        title = "{Stochastic Modeling of the Time Variability of ALMA Calibrators}",
      journal = {\pasp},
         year = 2019,
        month = sep,
       volume = {131},
       number = {1003},
        pages = {094504},
          doi = {10.1088/1538-3873/ab2d38},
archivePrefix = {arXiv},
       eprint = {1907.03528},
 primaryClass = {astro-ph.IM},
       adsurl = {https://ui.adsabs.harvard.edu/abs/2019PASP..131i4504G}
}

@ARTICLE{Sandstrom23,
       author = {{Sandstrom}, Karin M. and {Chastenet}, J{\'e}r{\'e}my and {Sutter}, Jessica and {Leroy}, Adam K. and {Egorov}, Oleg V. and {Williams}, Thomas G. and {Bolatto}, Alberto D. and {Boquien}, M{\'e}d{\'e}ric and {Cao}, Yixian and {Dale}, Daniel A. and {Lee}, Janice C. and {Rosolowsky}, Erik and {Schinnerer}, Eva and {Barnes}, Ashley. T. and {Belfiore}, Francesco and {Bigiel}, F. and {Chevance}, M{\'e}lanie and {Grasha}, Kathryn and {Groves}, Brent and {Hassani}, Hamid and {Hughes}, Annie and {Klessen}, Ralf S. and {Kruijssen}, J.~M. Diederik and {Larson}, Kirsten L. and {Liu}, Daizhong and {Lopez}, Laura A. and {Meidt}, Sharon E. and {Murphy}, Eric J. and {Sormani}, Mattia C. and {Thilker}, David A. and {Watkins}, Elizabeth J.},
        title = "{PHANGS-JWST First Results: Mapping the 3.3 {\ensuremath{\mu}}m Polycyclic Aromatic Hydrocarbon Vibrational Band in Nearby Galaxies with NIRCam Medium Bands}",
      journal = {\apjl},
         year = 2023,
        month = feb,
       volume = {944},
       number = {2},
          eid = {L7},
        pages = {L7},
          doi = {10.3847/2041-8213/acb0cf},
archivePrefix = {arXiv},
       eprint = {2301.00854},
 primaryClass = {astro-ph.GA},
       adsurl = {https://ui.adsabs.harvard.edu/abs/2023ApJ...944L...7S}
}

@ARTICLE{Sidhu22,
       author = {{Sidhu}, Ameek and {Tielens}, A.~G.~G.~M. and {Peeters}, Els and {Cami}, Jan},
        title = "{Polycyclic Aromatic Hydrocarbon emission model in photodissociation regions - I. Application to the 3.3, 6.2, and 11.2 {\ensuremath{\mu}}m bands}",
      journal = {\mnras},
         year = 2022,
        month = jul,
       volume = {514},
       number = {1},
        pages = {342-369},
          doi = {10.1093/mnras/stac1255},
archivePrefix = {arXiv},
       eprint = {2205.03304},
 primaryClass = {astro-ph.GA},
       adsurl = {https://ui.adsabs.harvard.edu/abs/2022MNRAS.514..342S}
}

@ARTICLE{Rigopoulou21,
       author = {{Rigopoulou}, D. and {Barale}, M. and {Clary}, D.~C. and {Shan}, X. and {Alonso-Herrero}, A. and {Garc{\'\i}a-Bernete}, I. and {Hunt}, L. and {Kerkeni}, B. and {Pereira-Santaella}, M. and {Roche}, P.~F.},
        title = "{The properties of polycyclic aromatic hydrocarbons in galaxies: constraints on PAH sizes, charge and radiation fields}",
      journal = {\mnras},
         year = 2021,
        month = jul,
       volume = {504},
       number = {4},
        pages = {5287-5300},
          doi = {10.1093/mnras/stab959},
archivePrefix = {arXiv},
       eprint = {2011.10114},
 primaryClass = {astro-ph.GA},
       adsurl = {https://ui.adsabs.harvard.edu/abs/2021MNRAS.504.5287R}
}

@ARTICLE{Smith07,
       author = {{Smith}, J.~D.~T. and {Draine}, B.~T. and {Dale}, D.~A. and {Moustakas}, J. and {Kennicutt}, Jr., R.~C. and {Helou}, G. and {Armus}, L. and {Roussel}, H. and {Sheth}, K. and {Bendo}, G.~J. and {Buckalew}, B.~A. and {Calzetti}, D. and {Engelbracht}, C.~W. and {Gordon}, K.~D. and {Hollenbach}, D.~J. and {Li}, A. and {Malhotra}, S. and {Murphy}, E.~J. and {Walter}, F.},
        title = "{The Mid-Infrared Spectrum of Star-forming Galaxies: Global Properties of Polycyclic Aromatic Hydrocarbon Emission}",
      journal = {\apj},
         year = 2007,
        month = feb,
       volume = {656},
       number = {2},
        pages = {770-791},
          doi = {10.1086/510549},
archivePrefix = {arXiv},
       eprint = {astro-ph/0610913},
 primaryClass = {astro-ph},
       adsurl = {https://ui.adsabs.harvard.edu/abs/2007ApJ...656..770S}
}

@ARTICLE{Sajina22,
       author = {{Sajina}, Anna and {Lacy}, Mark and {Pope}, Alexandra},
        title = "{The Past and Future of Mid-Infrared Studies of AGN}",
      journal = {Universe},
         year = 2022,
        month = jun,
       volume = {8},
       number = {7},
          eid = {356},
        pages = {356},
          doi = {10.3390/universe8070356},
archivePrefix = {arXiv},
       eprint = {2210.02307},
 primaryClass = {astro-ph.GA},
       adsurl = {https://ui.adsabs.harvard.edu/abs/2022Univ....8..356S}
}

@ARTICLE{Hankla25,
       author = {{Hankla}, A. and {Philippov}, A. and {Mbarek}, R. and {Mushotzky}, R. and {Musoke}, G. and {Gro{\v{s}}elj}, D. and {Liska}, M.},
        title = "{An outflow from the X-ray corona as the origin of millimeter emission from radio-quiet AGN}",
      journal = {arXiv e-prints},
         year = 2025,
        month = dec,
          eid = {arXiv:2512.01662},
        pages = {arXiv:2512.01662},
          doi = {10.48550/arXiv.2512.01662},
archivePrefix = {arXiv},
       eprint = {2512.01662},
 primaryClass = {astro-ph.HE},
       adsurl = {https://ui.adsabs.harvard.edu/abs/2025arXiv251201662H}
}

@ARTICLE{Yamada24,
       author = {{Yamada}, Tomoya and {Sakai}, Nobuyuki and {Inoue}, Yoshiyuki and {Michiyama}, Tomonari},
        title = "{Deciphering Radio Emissions from Accretion Disk Winds in Radio-quiet Active Galactic Nuclei}",
      journal = {\apj},
         year = 2024,
        month = jun,
       volume = {968},
       number = {2},
          eid = {116},
        pages = {116},
          doi = {10.3847/1538-4357/ad3a63},
archivePrefix = {arXiv},
       eprint = {2404.04632},
 primaryClass = {astro-ph.HE},
       adsurl = {https://ui.adsabs.harvard.edu/abs/2024ApJ...968..116Y}
}

@ARTICLE{Gallo19,
       author = {{Gallo}, Elena and {Teague}, Richard and {Plotkin}, Richard M. and {Miller-Jones}, James C.~A. and {Russell}, David M. and {Din{\c{c}}er}, Tolga and {Bailyn}, Charles and {Maccarone}, Thomas J. and {Markoff}, Sera and {Fender}, Rob P.},
        title = "{ALMA observations of A0620-00: fresh clues on the nature of quiescent black hole X-ray binary jets}",
      journal = {\mnras},
         year = 2019,
        month = sep,
       volume = {488},
       number = {1},
        pages = {191-197},
          doi = {10.1093/mnras/stz1634},
archivePrefix = {arXiv},
       eprint = {1906.04299},
 primaryClass = {astro-ph.HE},
       adsurl = {https://ui.adsabs.harvard.edu/abs/2019MNRAS.488..191G}
}

@ARTICLE{DiazTrigo21,
       author = {{D{\'\i}az Trigo}, M. and {Petry}, D. and {Humphreys}, E. and {Impellizzeri}, C.~M.~V. and {Liu}, H.~B.},
        title = "{A search for signatures of interactions of X-ray binary outflows with their environments with ALMA}",
      journal = {\aap},
         year = 2021,
        month = jun,
       volume = {650},
          eid = {A37},
        pages = {A37},
          doi = {10.1051/0004-6361/202040160},
archivePrefix = {arXiv},
       eprint = {2104.05384},
 primaryClass = {astro-ph.HE},
       adsurl = {https://ui.adsabs.harvard.edu/abs/2021A&A...650A..37D}
}

@ARTICLE{Shablovinskaya2024,
       author = {{Shablovinskaya}, E. and {Ricci}, C. and {Chang}, C. -S. and {Tortosa}, A. and {del Palacio}, S. and {Kawamuro}, T. and {Aalto}, S. and {Arzoumanian}, Z. and {Balokovic}, M. and {Bauer}, F.~E. and {Gendreau}, K.~C. and {Ho}, L.~C. and {Kakkad}, D. and {Kara}, E. and {Koss}, M.~J. and {Liu}, T. and {Loewenstein}, M. and {Mushotzky}, R. and {Paltani}, S. and {Privon}, G.~C. and {Smith}, K. and {Tombesi}, F. and {Trakhtenbrot}, B.},
        title = "{Joint ALMA/X-ray monitoring of the radio-quiet type 1 active galactic nucleus IC 4329A}",
      journal = {\aap},
         year = 2024,
        month = oct,
       volume = {690},
          eid = {A232},
        pages = {A232},
          doi = {10.1051/0004-6361/202450133},
archivePrefix = {arXiv},
       eprint = {2403.19524},
 primaryClass = {astro-ph.HE},
       adsurl = {https://ui.adsabs.harvard.edu/abs/2024A&A...690A.232S}
}

@ARTICLE{Michiyama2024,
       author = {{Michiyama}, Tomonari and {Inoue}, Yoshiyuki and {Doi}, Akihiro and {Yamada}, Tomoya and {Fukazawa}, Yasushi and {Kubo}, Hidetoshi and {Barnier}, Samuel},
        title = "{ALMA Confirmation of Millimeter Time Variability in the Gamma-Ray Detected Seyfert Galaxy GRS 1734-292}",
      journal = {\apj},
         year = 2024,
        month = apr,
       volume = {965},
       number = {1},
          eid = {68},
        pages = {68},
          doi = {10.3847/1538-4357/ad2fae},
archivePrefix = {arXiv},
       eprint = {2404.00647},
 primaryClass = {astro-ph.GA},
       adsurl = {https://ui.adsabs.harvard.edu/abs/2024ApJ...965...68M}
}

@ARTICLE{Inoue2018,
       author = {{Inoue}, Yoshiyuki and {Doi}, Akihiro},
        title = "{Detection of Coronal Magnetic Activity in nearby Active Supermassive Black Holes}",
      journal = {\apj},
         year = 2018,
        month = dec,
       volume = {869},
       number = {2},
          eid = {114},
        pages = {114},
          doi = {10.3847/1538-4357/aaeb95},
archivePrefix = {arXiv},
       eprint = {1810.10732},
 primaryClass = {astro-ph.HE},
       adsurl = {https://ui.adsabs.harvard.edu/abs/2018ApJ...869..114I}
}

@ARTICLE{Inoue2014,
       author = {{Inoue}, Yoshiyuki and {Doi}, Akihiro},
        title = "{Unveiling the nature of coronae in active galactic nuclei through submillimeter observations}",
      journal = {\pasj},
         year = 2014,
        month = dec,
       volume = {66},
       number = {6},
          eid = {L8},
        pages = {L8},
          doi = {10.1093/pasj/psu079},
archivePrefix = {arXiv},
       eprint = {1411.2334},
 primaryClass = {astro-ph.HE},
       adsurl = {https://ui.adsabs.harvard.edu/abs/2014PASJ...66L...8I}
}

@ARTICLE{Brandl2009,
       author = {{Brandl}, B.~R. and {Snijders}, L. and {den Brok}, M. and {Whelan}, D.~G. and {Groves}, B. and {van der Werf}, P. and {Charmandaris}, V. and {Smith}, J.~D. and {Armus}, L. and {Kennicutt}, Jr., R.~C. and {Houck}, J.~R.},
        title = "{Spitzer-IRS Study of the Antennae Galaxies NGC 4038/39}",
      journal = {\apj},
         year = 2009,
        month = jul,
       volume = {699},
       number = {2},
        pages = {1982-2001},
          doi = {10.1088/0004-637X/699/2/1982},
archivePrefix = {arXiv},
       eprint = {0905.1058},
 primaryClass = {astro-ph.CO},
       adsurl = {https://ui.adsabs.harvard.edu/abs/2009ApJ...699.1982B}
}

@ARTICLE{Gilbert2000,
       author = {{Gilbert}, Andrea M. and {Graham}, James R. and {McLean}, Ian S. and {Becklin}, E.~E. and {Figer}, Donald F. and {Larkin}, James E. and {Levenson}, N.~A. and {Teplitz}, Harry I. and {Wilcox}, Mavourneen K.},
        title = "{Infrared Spectroscopy of a Massive Obscured Star Cluster in the Antennae Galaxies (NGC 4038/9) with NIRSPEC}",
      journal = {\apjl},
         year = 2000,
        month = apr,
       volume = {533},
       number = {1},
        pages = {L57-L60},
          doi = {10.1086/312599},
archivePrefix = {arXiv},
       eprint = {astro-ph/9912369},
 primaryClass = {astro-ph},
       adsurl = {https://ui.adsabs.harvard.edu/abs/2000ApJ...533L..57G}
}

@ARTICLE{Zezas2002a,
       author = {{Zezas}, A. and {Fabbiano}, G. and {Rots}, A.~H. and {Murray}, S.~S.},
        title = "{Chandra Observations of ``The Antennae'' Galaxies (NGC 4038/4039). II. Detection and Analysis of Galaxian X-Ray Sources}",
      journal = {\apjs},
         year = 2002,
        month = oct,
       volume = {142},
       number = {2},
        pages = {239-260},
          doi = {10.1086/342010},
archivePrefix = {arXiv},
       eprint = {astro-ph/0203174},
 primaryClass = {astro-ph},
       adsurl = {https://ui.adsabs.harvard.edu/abs/2002ApJS..142..239Z}
}

@ARTICLE{Zezas2002b,
       author = {{Zezas}, A. and {Fabbiano}, G. and {Rots}, A.~H. and {Murray}, S.~S.},
        title = "{Chandra Observations of ``The Antennae'' Galaxies (NGC 4038/4039). III. X-Ray Properties and Multiwavelength Associations of the X-Ray Source Population}",
      journal = {\apj},
         year = 2002,
        month = oct,
       volume = {577},
       number = {2},
        pages = {710-725},
          doi = {10.1086/342160},
archivePrefix = {arXiv},
       eprint = {astro-ph/0203175},
 primaryClass = {astro-ph},
       adsurl = {https://ui.adsabs.harvard.edu/abs/2002ApJ...577..710Z}
}

@ARTICLE{Neff2000,
       author = {{Neff}, Susan G. and {Ulvestad}, James S.},
        title = "{VLA Observations of the Nearby Merger NGC 4038/4039: H II Regions and Supernova Remnants in the ``Antennae''}",
      journal = {\aj},
         year = 2000,
        month = aug,
       volume = {120},
       number = {2},
        pages = {670-696},
          doi = {10.1086/301503},
       adsurl = {https://ui.adsabs.harvard.edu/abs/2000AJ....120..670N}
}

@ARTICLE{He2022,
       author = {{He}, Hao and {Wilson}, Christine and {Brunetti}, Nathan and {Finn}, Molly and {Bemis}, Ashley and {Johnson}, Kelsey},
        title = "{Embedded Young Massive Star Clusters in the Antennae Merger}",
      journal = {\apj},
         year = 2022,
        month = mar,
       volume = {928},
       number = {1},
          eid = {57},
        pages = {57},
          doi = {10.3847/1538-4357/ac5628},
archivePrefix = {arXiv},
       eprint = {2202.08077},
 primaryClass = {astro-ph.GA},
       adsurl = {https://ui.adsabs.harvard.edu/abs/2022ApJ...928...57H}
}

@ARTICLE{Poutanen2013,
       author = {{Poutanen}, Juri and {Fabrika}, Sergei and {Valeev}, Azamat F. and {Sholukhova}, Olga and {Greiner}, Jochen},
        title = "{On the association of the ultraluminous X-ray sources in the Antennae galaxies with young stellar clusters}",
      journal = {\mnras},
         year = 2013,
        month = jun,
       volume = {432},
       number = {1},
        pages = {506-519},
          doi = {10.1093/mnras/stt487},
archivePrefix = {arXiv},
       eprint = {1210.1210},
 primaryClass = {astro-ph.HE},
       adsurl = {https://ui.adsabs.harvard.edu/abs/2013MNRAS.432..506P}
}

@ARTICLE{Ricci2023,
       author = {{Ricci}, Claudio and {Chang}, Chin-Shin and {Kawamuro}, Taiki and {Privon}, George C. and {Mushotzky}, Richard and {Trakhtenbrot}, Benny and {Laor}, Ari and {Koss}, Michael J. and {Smith}, Krista L. and {Gupta}, Kriti K. and {Dimopoulos}, Georgios and {Aalto}, Susanne and {Ros}, Eduardo},
        title = "{A Tight Correlation between Millimeter and X-Ray Emission in Accreting Massive Black Holes from <100 mas Resolution ALMA Observations}",
      journal = {\apjl},
         year = 2023,
        month = aug,
       volume = {952},
       number = {2},
          eid = {L28},
        pages = {L28},
          doi = {10.3847/2041-8213/acda27},
archivePrefix = {arXiv},
       eprint = {2306.04679},
 primaryClass = {astro-ph.HE},
       adsurl = {https://ui.adsabs.harvard.edu/abs/2023ApJ...952L..28R}
}

@ARTICLE{Oh2018,
       author = {{Oh}, Kyuseok and {Koss}, Michael and {Markwardt}, Craig B. and {Schawinski}, Kevin and {Baumgartner}, Wayne H. and {Barthelmy}, Scott D. and {Cenko}, S. Bradley and {Gehrels}, Neil and {Mushotzky}, Richard and {Petulante}, Abigail and {Ricci}, Claudio and {Lien}, Amy and {Trakhtenbrot}, Benny},
        title = "{The 105-Month Swift-BAT All-sky Hard X-Ray Survey}",
      journal = {\apjs},
         year = 2018,
        month = mar,
       volume = {235},
       number = {1},
          eid = {4},
        pages = {4},
          doi = {10.3847/1538-4365/aaa7fd},
archivePrefix = {arXiv},
       eprint = {1801.01882},
 primaryClass = {astro-ph.HE},
       adsurl = {https://ui.adsabs.harvard.edu/abs/2018ApJS..235....4O}
}

@ARTICLE{Chastenet2023,
       author = {{Chastenet}, J{\'e}r{\'e}my and {Sutter}, Jessica and {Sandstrom}, Karin and {Belfiore}, Francesco and {Egorov}, Oleg V. and {Larson}, Kirsten L. and {Leroy}, Adam K. and {Liu}, Daizhong and {Rosolowsky}, Erik and {Thilker}, David A. and {Watkins}, Elizabeth J. and {Williams}, Thomas G. and {Barnes}, Ashley. T. and {Bigiel}, F. and {Boquien}, M{\'e}d{\'e}ric and {Chevance}, M{\'e}lanie and {Dale}, Daniel A. and {Kruijssen}, J.~M. Diederik and {Emsellem}, Eric and {Grasha}, Kathryn and {Groves}, Brent and {Hassani}, Hamid and {Hughes}, Annie and {Kreckel}, Kathryn and {Meidt}, Sharon E. and {Pan}, Hsi-An and {Querejeta}, Miguel and {Schinnerer}, Eva and {Whitcomb}, Cory M.},
        title = "{PHANGS-JWST First Results: Measuring Polycyclic Aromatic Hydrocarbon Properties across the Multiphase Interstellar Medium}",
      journal = {\apjl},
         year = 2023,
        month = feb,
       volume = {944},
       number = {2},
          eid = {L12},
        pages = {L12},
          doi = {10.3847/2041-8213/acac94},
       adsurl = {https://ui.adsabs.harvard.edu/abs/2023ApJ...944L..12C}
}

@ARTICLE{Zhang2025,
       author = {{Zhang}, Congcong and {Hales}, Joelene and {Peeters}, Els and {Cami}, Jan and {Sidhu}, Ameek and {Zhen}, Junfeng},
        title = "{A JWST Study of Polycyclic Aromatic Hydrocarbon Emission in a Region of 30 Doradus}",
      journal = {\apjs},
         year = 2025,
        month = sep,
       volume = {280},
       number = {1},
          eid = {4},
        pages = {4},
          doi = {10.3847/1538-4365/adea6b},
archivePrefix = {arXiv},
       eprint = {2410.18909},
 primaryClass = {astro-ph.GA},
       adsurl = {https://ui.adsabs.harvard.edu/abs/2025ApJS..280....4Z}
}

@ARTICLE{Li2023,
       author = {{Li}, Wenhao and {Nair}, Preethi and {Irwin}, Jimmy and {Ellison}, Sara and {Satyapal}, Shobita and {Drory}, Niv and {Jones}, Amy and {Keel}, William and {Masters}, Karen and {Stark}, David and {Ryan}, Russell and {Mukundan}, Kavya},
        title = "{A Multiwavelength Study of Active Galactic Nuclei in Post-merger Remnants}",
      journal = {\apj},
         year = 2023,
        month = feb,
       volume = {944},
       number = {2},
          eid = {168},
        pages = {168},
          doi = {10.3847/1538-4357/acb13d},
archivePrefix = {arXiv},
       eprint = {2301.06186},
 primaryClass = {astro-ph.GA},
       adsurl = {https://ui.adsabs.harvard.edu/abs/2023ApJ...944..168L}
}

@ARTICLE{Renaud2009,
       author = {{Renaud}, F. and {Boily}, C.~M. and {Naab}, T. and {Theis}, Ch.},
        title = "{Fully Compressive Tides in Galaxy Mergers}",
      journal = {\apj},
         year = 2009,
        month = nov,
       volume = {706},
       number = {1},
        pages = {67-82},
          doi = {10.1088/0004-637X/706/1/67},
archivePrefix = {arXiv},
       eprint = {0910.0196},
 primaryClass = {astro-ph.CO},
       adsurl = {https://ui.adsabs.harvard.edu/abs/2009ApJ...706...67R}
}

@ARTICLE{Kawamuro2022,
       author = {{Kawamuro}, Taiki and {Ricci}, Claudio and {Imanishi}, Masatoshi and {Mushotzky}, Richard F. and {Izumi}, Takuma and {Ricci}, Federica and {Bauer}, Franz E. and {Koss}, Michael J. and {Trakhtenbrot}, Benny and {Ichikawa}, Kohei and {Rojas}, Alejandra F. and {Smith}, Krista Lynne and {Shimizu}, Taro and {Oh}, Kyuseok and {den Brok}, Jakob S. and {Baba}, Shunsuke and {Balokovi{\'c}}, Mislav and {Chang}, Chin-Shin and {Kakkad}, Darshan and {Pfeifle}, Ryan W. and {Privon}, George C. and {Temple}, Matthew J. and {Ueda}, Yoshihiro and {Harrison}, Fiona and {Powell}, Meredith C. and {Stern}, Daniel and {Urry}, Meg and {Sanders}, David B.},
        title = "{BASS XXXII: Studying the Nuclear Millimeter-wave Continuum Emission of AGNs with ALMA at Scales {\ensuremath{\lesssim}}100-200 pc}",
      journal = {\apj},
         year = 2022,
        month = oct,
       volume = {938},
       number = {1},
          eid = {87},
        pages = {87},
          doi = {10.3847/1538-4357/ac8794},
archivePrefix = {arXiv},
       eprint = {2208.03880},
 primaryClass = {astro-ph.GA},
       adsurl = {https://ui.adsabs.harvard.edu/abs/2022ApJ...938...87K}
}

@ARTICLE{Schweizer08,
       author = {{Schweizer}, Fran{\c{c}}ois and {Burns}, Christopher R. and {Madore}, Barry F. and {Mager}, Violet A. and {Phillips}, M.~M. and {Freedman}, Wendy L. and {Boldt}, Luis and {Contreras}, Carlos and {Folatelli}, Gaston and {Gonz{\'a}lez}, Sergio and {Hamuy}, Mario and {Krzeminski}, Wojtek and {Morrell}, Nidia I. and {Persson}, S.~E. and {Roth}, Miguel R. and {Stritzinger}, Maximilian D.},
        title = "{A New Distance to the Antennae Galaxies (NGC 4038/39) Based on the Type Ia Supernova 2007sr}",
      journal = {\aj},
         year = 2008,
        month = oct,
       volume = {136},
       number = {4},
        pages = {1482-1489},
          doi = {10.1088/0004-6256/136/4/1482},
archivePrefix = {arXiv},
       eprint = {0807.3955},
 primaryClass = {astro-ph},
       adsurl = {https://ui.adsabs.harvard.edu/abs/2008AJ....136.1482S}
}

@ARTICLE{CASA22,
       author = {{CASA Team} and {Bean}, Ben and {Bhatnagar}, Sanjay and {Castro}, Sandra and {Donovan Meyer}, Jennifer and {Emonts}, Bjorn and {Garcia}, Enrique and {Garwood}, Robert and {Golap}, Kumar and {Gonzalez Villalba}, Justo and {Harris}, Pamela and {Hayashi}, Yohei and {Hoskins}, Josh and {Hsieh}, Mingyu and {Jagannathan}, Preshanth and {Kawasaki}, Wataru and {Keimpema}, Aard and {Kettenis}, Mark and {Lopez}, Jorge and {Marvil}, Joshua and {Masters}, Joseph and {McNichols}, Andrew and {Mehringer}, David and {Miel}, Renaud and {Moellenbrock}, George and {Montesino}, Federico and {Nakazato}, Takeshi and {Ott}, Juergen and {Petry}, Dirk and {Pokorny}, Martin and {Raba}, Ryan and {Rau}, Urvashi and {Schiebel}, Darrell and {Schweighart}, Neal and {Sekhar}, Srikrishna and {Shimada}, Kazuhiko and {Small}, Des and {Steeb}, Jan-Willem and {Sugimoto}, Kanako and {Suoranta}, Ville and {Tsutsumi}, Takahiro and {van Bemmel}, Ilse M. and {Verkouter}, Marjolein and {Wells}, Akeem and {Xiong}, Wei and {Szomoru}, Arpad and {Griffith}, Morgan and {Glendenning}, Brian and {Kern}, Jeff},
        title = "{CASA, the Common Astronomy Software Applications for Radio Astronomy}",
      journal = {\pasp},
         year = 2022,
        month = nov,
       volume = {134},
       number = {1041},
          eid = {114501},
        pages = {114501},
          doi = {10.1088/1538-3873/ac9642},
archivePrefix = {arXiv},
       eprint = {2210.02276},
 primaryClass = {astro-ph.IM},
       adsurl = {https://ui.adsabs.harvard.edu/abs/2022PASP..134k4501C}
}

@ARTICLE{Hunter23,
       author = {{Hunter}, Todd R. and {Indebetouw}, Remy and {Brogan}, Crystal L. and {Berry}, Kristin and {Chang}, Chin-Shin and {Francke}, Harold and {Geers}, Vincent C. and {G{\'o}mez}, Laura and {Hibbard}, John E. and {Humphreys}, Elizabeth M. and {Kent}, Brian R. and {Kepley}, Amanda A. and {Kunneriath}, Devaky and {Lipnicky}, Andrew and {Loomis}, Ryan A. and {Mason}, Brian S. and {Masters}, Joseph S. and {Maud}, Luke T. and {Muders}, Dirk and {Sabater}, Jose and {Sugimoto}, Kanako and {Sz{\H{u}}cs}, L{\'a}szl{\'o} and {Vasiliev}, Eugene and {Videla}, Liza and {Villard}, Eric and {Williams}, Stewart J. and {Xue}, Rui and {Yoon}, Ilsang},
        title = "{The ALMA Interferometric Pipeline Heuristics}",
      journal = {\pasp},
         year = 2023,
        month = jul,
       volume = {135},
       number = {1049},
          eid = {074501},
        pages = {074501},
          doi = {10.1088/1538-3873/ace216},
archivePrefix = {arXiv},
       eprint = {2306.07420},
 primaryClass = {astro-ph.IM},
       adsurl = {https://ui.adsabs.harvard.edu/abs/2023PASP..135g4501H}
}

@ARTICLE{Chomiuk09,
       author = {{Chomiuk}, Laura and {Wilcots}, Eric M.},
        title = "{A Universal Luminosity Function for Radio Supernova Remnants}",
      journal = {\apj},
         year = 2009,
        month = sep,
       volume = {703},
       number = {1},
        pages = {370-389},
          doi = {10.1088/0004-637X/703/1/370},
archivePrefix = {arXiv},
       eprint = {0907.4783},
 primaryClass = {astro-ph.CO},
       adsurl = {https://ui.adsabs.harvard.edu/abs/2009ApJ...703..370C}
}

@ARTICLE{Cseh2012,
       author = {{Cseh}, D{\'a}vid and {Corbel}, St{\'e}phane and {Kaaret}, Philip and {Lang}, Cornelia and {Gris{\'e}}, Fabien and {Paragi}, Zsolt and {Tzioumis}, Anastasios and {Tudose}, Valeriu and {Feng}, Hua},
        title = "{Black Hole Powered Nebulae and a Case Study of the Ultraluminous X-Ray Source IC 342 X-1}",
      journal = {\apj},
         year = 2012,
        month = apr,
       volume = {749},
       number = {1},
          eid = {17},
        pages = {17},
          doi = {10.1088/0004-637X/749/1/17},
archivePrefix = {arXiv},
       eprint = {1201.4473},
 primaryClass = {astro-ph.HE},
       adsurl = {https://ui.adsabs.harvard.edu/abs/2012ApJ...749...17C}
}

@ARTICLE{Marti2018,
       author = {{Mart{\'\i}}, Josep and {Bujalance-Fern{\'a}ndez}, Irene and {Luque-Escamilla}, Pedro L. and {S{\'a}nchez-Ayaso}, Estrella and {Paredes}, Josep M. and {Rib{\'o}}, Marc},
        title = "{The radio jets of SS 433 at millimetre wavelengths}",
      journal = {\aap},
         year = 2018,
        month = nov,
       volume = {619},
          eid = {A40},
        pages = {A40},
          doi = {10.1051/0004-6361/201833733},
archivePrefix = {arXiv},
       eprint = {1809.01386},
 primaryClass = {astro-ph.HE},
       adsurl = {https://ui.adsabs.harvard.edu/abs/2018A&A...619A..40M}
}

@ARTICLE{Lockman2007,
       author = {{Lockman}, Felix J. and {Blundell}, Katherine M. and {Goss}, W.~M.},
        title = "{The distance to SS433/W50 and its interaction with the interstellar medium}",
      journal = {\mnras},
         year = 2007,
        month = nov,
       volume = {381},
       number = {3},
        pages = {881-893},
          doi = {10.1111/j.1365-2966.2007.12170.x},
archivePrefix = {arXiv},
       eprint = {0707.0506},
 primaryClass = {astro-ph},
       adsurl = {https://ui.adsabs.harvard.edu/abs/2007MNRAS.381..881L}
}

@ARTICLE{Koljonen2021,
       author = {{Koljonen}, K.~I.~I. and {Hovatta}, T.},
        title = "{ALMA/NICER observations of GRS 1915+105 indicate a return to a hard state}",
      journal = {\aap},
         year = 2021,
        month = mar,
       volume = {647},
          eid = {A173},
        pages = {A173},
          doi = {10.1051/0004-6361/202039581},
archivePrefix = {arXiv},
       eprint = {2102.00693},
 primaryClass = {astro-ph.HE},
       adsurl = {https://ui.adsabs.harvard.edu/abs/2021A&A...647A.173K}
}

@ARTICLE{Fender1999,
       author = {{Fender}, R.~P. and {Garrington}, S.~T. and {McKay}, D.~J. and {Muxlow}, T.~W.~B. and {Pooley}, G.~G. and {Spencer}, R.~E. and {Stirling}, A.~M. and {Waltman}, E.~B.},
        title = "{MERLIN observations of relativistic ejections from GRS 1915+105}",
      journal = {\mnras},
         year = 1999,
        month = apr,
       volume = {304},
       number = {4},
        pages = {865-876},
          doi = {10.1046/j.1365-8711.1999.02364.x},
archivePrefix = {arXiv},
       eprint = {astro-ph/9812150},
 primaryClass = {astro-ph},
       adsurl = {https://ui.adsabs.harvard.edu/abs/1999MNRAS.304..865F}
}

@ARTICLE{CRC2,
       author = {{Evans}, Ian N. and {Evans}, Janet D. and {Mart{\'\i}nez-Galarza}, J. Rafael and {Miller}, Joseph B. and {Primini}, Francis A. and {Azadi}, Mojegan and {Burke}, Douglas J. and {Civano}, Francesca M. and {D'Abrusco}, Raffaele and {Fabbiano}, Giuseppina and {Graessle}, Dale E. and {Grier}, John D. and {Houck}, John C. and {Lauer}, Jennifer and {McCollough}, Michael L. and {Nowak}, Michael A. and {Plummer}, David A. and {Rots}, Arnold H. and {Siemiginowska}, Aneta and {Tibbetts}, Michael S.},
        title = "{The Chandra Source Catalog Release 2 Series}",
      journal = {\apjs},
         year = 2024,
        month = oct,
       volume = {274},
       number = {2},
          eid = {22},
        pages = {22},
          doi = {10.3847/1538-4365/ad6319},
archivePrefix = {arXiv},
       eprint = {2407.10799},
 primaryClass = {astro-ph.HE},
       adsurl = {https://ui.adsabs.harvard.edu/abs/2024ApJS..274...22E}
}
\bibliographystyle{aasjournal}

\end{document}